\documentclass[journal,comsoc]{IEEEtran}

\usepackage{amsmath, graphics,amssymb,epsfig,subfigure,color,amsthm,cite}
\usepackage{array}
\usepackage{multirow}
\usepackage{enumerate}
\usepackage{algorithmic}
\usepackage{algorithm}
\usepackage{bm}
\usepackage{algorithmic}
\usepackage[caption=false]{subfig}

\newtheorem{thm}{Theorem}

\newtheorem{cor}{Corollary}

\begin{document}
\title{Supervised-Learning-Aided Communication Framework for MIMO Systems with Low-Resolution ADCs}

\author{Yo-Seb Jeon, Song-Nam Hong, and Namyoon Lee
\thanks{Y.-S. Jeon and N. Lee are with the Department of Electrical Engineering, POSTECH, Pohang, Gyeongbuk 37673, South Korea (e-mail: yoseb.jeon@postech.ac.kr, nylee@postech.ac.kr).}
\thanks{S.-N. Hong is with the Department of Electrical and Computer Engineering, Ajou University, Suwon, Gyeonggi 16499, South Korea (e-mail: snhong@ajou.ac.kr).}}
\vspace{-2mm}

\maketitle

\newcommand{\argmax}{\operatornamewithlimits{argmax}}
\newcommand{\argmin}{\operatornamewithlimits{argmin}}
\makeatletter
\newcommand{\vast}{\bBigg@{3.2}}
\newcommand{\Vast}{\bBigg@{4.5}}
\makeatother
\vspace{-12mm}

\begin{abstract} 
This paper considers a multiple-input-multiple-output (MIMO) system with low-resolution analog-to-digital converters (ADCs). In this system, we propose a novel communication framework that is inspired by supervised learning. The key idea of the proposed framework is to learn the non-linear input-output system, formed by the concatenation of a wireless channel and a quantization function used at the ADCs, for data detection. In this framework, a conventional channel estimation process is replaced by a system learning process, in which the conditional probability mass functions (PMFs) of the nonlinear system are empirically learned by sending the repetitions of all possible data signals as pilot signals. Then the subsequent data detection process is performed based on the empirical conditional PMFs obtained during the system learning. To reduce both the training overhead and the detection complexity, we also develop a supervised-learning-aided successive-interference-cancellation method. In this method, a data signal vector is divided into two subvectors with reduced dimensions. Then these two subvectors are successively detected based on the conditional PMFs that are learned using artificial noise signals and an estimated channel. For the case of one-bit ADCs, we derive an analytical expression for vector-error-rate of the proposed framework under perfect channel knowledge at the receiver. Simulations demonstrate the detection error reduction of the proposed framework compared to conventional detection techniques that are based on channel estimation.
\end{abstract}

\begin{IEEEkeywords}
Multiple-input-multiple-output (MIMO) detection, data detection, one-bit analog-to-digital converter (ADC), massive MIMO, supervised learning.
\end{IEEEkeywords}

\section{Introduction}
\IEEEPARstart{F}{uture} wireless systems are possible to provide communication links with Gbps data rates by using a massive antenna array and/or by using a wide (possibly multi-gigahertz) bandwidth \cite{Pi2011,Swindlehurst2014,Patcharam2016,Lin2017}. The use of massive number of antennas and a wide-bandwidth causes significant power consumption at a receiver because of high-resolution (e.g., $8{\sim}14$-bit precision) analog-to-digital converters (ADCs). For example, the power consumption of the ADCs is shown to increase with both the number of precision levels and the system bandwidth (i.e., the  Nyquist sampling rate) \cite{Murmann,Walden1999,MezghaniNossek2010}. Therefore, the use of low-resolution (e.g., $1{\sim}3$-bit precision) ADCs has been regarded as a cost-effective solution to reduce the power consumption of future wireless systems including massive multiple-input-multiple-output (MIMO) systems and wideband communication systems \cite{Nossek2006,MezghaniNossek2007,Madhow2009,Mo2015,Bjornson2015,Mollen2017,Dong2017}. Unfortunately, when employing the low-resolution ADCs, a conventional linear signal model is changed into a nonlinear model due to the coarse quantization effect by the ADCs. Therefore, in this case, conventional data detection methods that ignore the quantization effect suffer from a significant performance loss.

Numerous detection methods have been proposed for MIMO systems with low-resolution ADCs, in order to deal with the nonlinear signal model formed by the quantization function at the ADCs \cite{Wang2014,Studer2016,Choi2016,Jeon2018,Jacobsson2017,Mezghani2007,Hong2017}. The optimal maximum-likelihood detection (MLD) was introduced for frequency-flat channels \cite{Wang2014} and for frequency-selective channels \cite{Studer2016}. In \cite{Wang2014,Studer2016}, it was shown that the MLD for the MIMO systems with the low-resolution ADCs is no more equivalent to the minimum Euclidean-distance detection. Some low-complexity variations of the MLD were also developed in \cite{Choi2016,Jeon2018}. The common idea of these methods is to find a reduced search space for the MLD without causing a significant performance loss. Particularly in \cite{Jeon2018}, it was shown that the MLD for the MIMO systems with one-bit ADCs is closely approximated by a weighted minimum Hamming-distance detection. Linear-detection methods such as zero-forcing detection \cite{Jacobsson2017} and  minimum-mean-square-error (MMSE) detection \cite{Studer2016,Mezghani2007} were considered to provide more affordable detection complexities. Their performances, however, are severely limited compared to the MLD-like detection methods. Recently, the authors in \cite{Hong2017} proposed an interesting solution for the data detection problem in the MIMO systems with the low-resolution ADCs by using modulo-type ADCs and lattice coding theory.

Most existing MIMO detection techniques have been developed under the assumption of estimated or perfect channel-state-information at the receiver (CSIR), to perform coherent detection. In practical systems, CSIR is attained by a channel estimation process that uses pilot signals with finite length. For MIMO systems with low-resolution ADCs, several channel estimation methods have been developed to improve the accuracy of CSIR \cite{Choi2016,Dabeer2010,Risi2016,Li2017,Studer2016,Wen2016}. ML-based channel estimators were developed for one-bit ADCs \cite{Choi2016} and for multi-bit ADCs \cite{Studer2016}, by formulating a convex problem that can be solved by an iterative algorithm. Linear channel estimators were also developed by using a least-squares method \cite{Risi2016} and by using the Bussgang decomposition \cite{Li2017}. Recently, an iterative algorithm that jointly estimates channel and data signals was proposed in \cite{Wen2016} by applying generalized approximate massage passing (GAMP) based on Bayesian inference theory. 
Despite the above efforts, when the number of bit precisions is extremely low ($1{\sim}2$-bit precision), the accuracy of CSIR obtained by the existing methods is severely limited by the coarse quantization effect at the ADCs, as reported in \cite{Li2017,Studer2016,Wen2016}.

In this paper, we study a data detection problem in MIMO systems with low-resolution (e.g. 1$\sim$3-bit precision) ADCs. For these systems, we propose a novel communication framework inspired by supervised learning. The major contributions of this paper are summarized as follows:

\begin{itemize}
  \item We propose a supervised-learning-aided communication framework for data detection in a MIMO system with low-resolution ADCs. The key idea of the proposed framework is to learn the nonlinear input-output system, formed by the concatenation of a wireless channel and a quantization function used at the ADCs. The proposed framework consists of two phases: 1) system learning and 2) data detection. For the system learning phase, we develop two learning methods that empirically estimate the conditional probability mass functions (PMFs) by using the repetitions of all possible data signals as pilot signals. For the data detection phase, we develop two detection methods, referred to as empirical maximum-likelihood detection (eMLD) and minimum-center-distance detection (MCD), that exploit the empirical conditional PMFs obtained from the system learning phase for the data detection. One salient feature of the proposed framework is that it requires nor CSIR or the knowledge of the quantization function used at the ADCs. It is also shown that the proposed framework with eMLD approaches to the optimal MLD with perfect CSIR, as the number of the training repetitions goes to infinity.

  \item We also develop a supervised-learning-aided successive-interference-cancellation (SL-SIC) that reduces both the training overhead and the detection complexity of the proposed framework. The fundamental of SL-SIC is to divide a symbol vector into two subvectors with reduced dimensions, and then to detect these two subvectors successively using the proposed framework. The developed SL-SIC consists of three phases: 1) symbol vector division, 2) system learning, and 3) data detection. For the symbol vector division phase, we devise an algorithm that divides a transmit symbol vector into two subvectors so that the chordal distance between two channel subspaces, each associating with one subvector, is maximized. For the system learning phase, we develop a modified learning method that learns the input-output relation between the first subvector and the received quantized vector, while marginalizing the effect of the second subvector. For the data detection phase, we introduce a two-stage MCD method that estimates the first subvector based on the learning information and then estimates the second subvector using the estimated first subvector. The developed SL-SIC provides a better performance-complexity tradeoff than the supervised-learning-aided communication framework, particularly when the modulation size or the number of transmit antennas is large.

  \item We analyze the vector-error-rate (VER) of the supervised-learning-aided communication framework when employing the one-bit ADCs. To the best of the authors' knowledge, no prior work has provided the VER performance analysis, because such analysis is not trivial when one-bit ADCs are taken into account in the MIMO system. Our approach for the analysis is to treat all possible quantized vectors as codewords of a nonlinear error-correcting code. Using this approach, we derive an upper bound of the VER in a closed-form for a fixed channel matrix in terms of the minimum Hamming distance $d_{\rm min}$ of the code. One major observation is that the upper bound of the VER decreases exponentially with the inverse of the number of transmit antennas, SNR, the minimum effective channel gain, and the minimum distance $d_{\rm min}$ that can increase with the number of receive antennas. To provide a more clear understanding for the VER performance under a random channel realization, we derive the approximate distribution of $d_{\rm min}$ in a closed-form, assuming a Rayleigh-fading channel distribution and binary-phase-shift-keying (BPSK) modulation. In particular, for the case of $N_{\rm t}=2$, we provide an exact distribution of $d_{\rm min}$. Our analysis results show that $d_{\rm min}$ increases linearly with $N_{\rm r}$; this can be interpreted as a diversity gain in the MIMO system with low-resolution ADCs.

  \item Using simulations, we evaluate the symbol-error-rate (SER) performance of both the supervised-learning-aided communication framework and the developed SL-SIC compared to existing detection techniques for MIMO systems with low-resolution ADCs. Simulation results show that the supervised-learning-aided framework outperforms the existing techniques that are based on channel estimation, when employing the same pilot length. It is also shown that the developed SL-SIC provides a better tradeoff between the SER performance and the detection complexity than the existing techniques. Using simulations, we also show the validation of the analysis results.
\end{itemize}

The supervised-learning-aided communication framework was originally introduced in \cite{Jeon2017} by the authors of this paper. We extend this framework by developing a more efficient learning method than the original method, and also by developing the SL-SIC method which has not been considered in \cite{Jeon2017}. In addition, we also provide more rigorous analysis and simulation results for the proposed framework, compared to the original work. Recent studies that use machine learning theory for wireless communications can be found in \cite{Daniels2008,Daniels2010,Nachmani2016,Gruber2017}, including adaptive modulation and coding problem \cite{Daniels2008,Daniels2010}, and the decoding problem of channel code \cite{Nachmani2016,Gruber2017}.

\subsubsection*{Notation}
Upper-case and lower-case boldface letters denote matrices and column vectors, respectively.
$\mathbb{E}[\cdot]$ is the statistical expectation,
    $\mathbb{P}(\cdot)$ is the probability,
    $(\cdot)^\top$ is the transpose,
    $(\cdot)^H$ is the conjugate transpose,
    $|\cdot|$ is the absolute value,
    $\text{Re}(\cdot)$ is the real part,
    $\text{Im}(\cdot)$ is the imaginary part,
    and $\lfloor\cdot\rfloor$ is the floor function.
$\Phi(\cdot)$ is the cumulative distribution of the standard normal random variable.

\section{System Model}\label{sec:model}
In this section, we present a system model for a MIMO system with low-resolution ADCs.

\begin{figure}[t]
        \centering
        {\epsfig{file=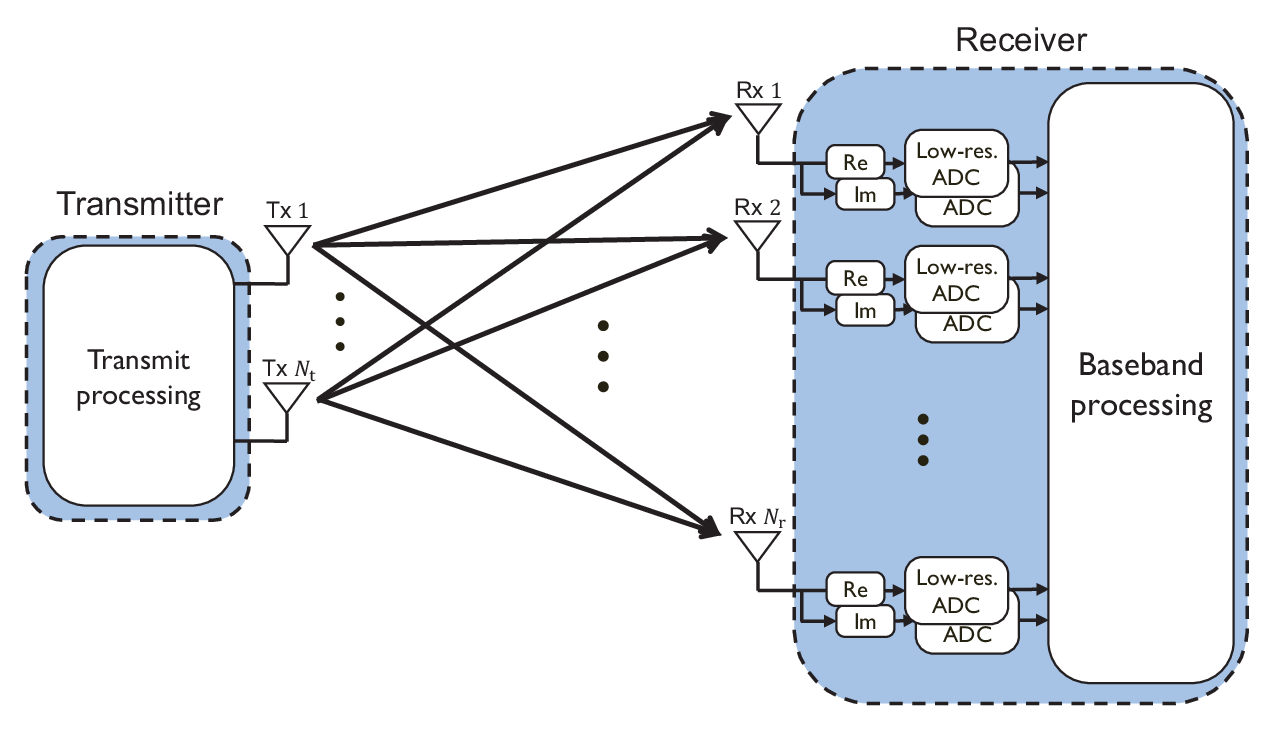, width=9cm}}
        \caption{Illustration of a MIMO system operating with low-resolution ADCs.}
        \label{fig:SM}
    \end{figure}

We consider a MIMO system with low-resolution ADCs as illustrated in Fig.~\ref{fig:SM}. In the considered system, a transmitter equipped with $N_{\rm t}$ transmit antennas intends to send $N_{\rm t}$ independent data symbols to a receiver equipped with $N_{\rm r}$ receive antennas. Let ${\bf x}[n] =[x_1[n], x_2[n],\ldots, x_{N_{\rm t}}[n]]^{\top}\in \mathbb{C}^{N_{\rm t}}$ be the data symbol vector sent by the transmitter at time slot $n$. Under the assumptions of the Nyquist sampling rate and perfect timing synchronization, the received signal vector ${\bf r}[n] \in \mathbb{C}^{N_{\rm r}}$ at time slot $n$ before the ADCs is
\begin{align}\label{Eq_rn}
    {\bf r}[n] = {\bf H}{\bf x}[n] + {\bf z}[n],
\end{align}
where ${\bf H}\in\mathbb{C}^{N_{\rm r}\times N_{\rm t}}$ denotes a channel matrix, and ${\bf z}[n]=[z_1[n], z_2[n],\ldots, z_{N_{\rm r}}[n]]^{\top}$ is a noise vector in which the elements are independent and identically distributed (i.i.d.) circularly-symmetric complex Gaussian random variables with zero mean and variance $\sigma^2$, i.e., $z_i[n]\sim\mathcal{CN}(0,\sigma^2)$. Each data symbol $x_i[n]$ satisfies $\mathbb{E}[|x_i[n]|^2]=1$ and is drawn from a constellation set $\mathcal{X}$ with constellation size $M=|\mathcal{X}|$. For instance,  $\mathcal{X}=\{-1,+1\}$ for BPSK modulation. The SNR of the considered system is defined as $\rho = \frac{N_{\rm t}}{\sigma^2}$.

We assume a block fading channel in which the channel remains constant for $T$ time slots. A transmission frame containing $T$ time slots consists of two different types of a frame: 1) a pilot transmission frame and 2) a data transmission frame. The first $T_{\rm t}$ time slots are allocated for the pilot transmission frame, and the subsequent $T_{\rm d}$ time slots are allocated for the data transmission frame, i.e., $T=T_{\rm t} + T_{\rm d}$. 

Each receive antenna is equipped with two low-resolution ADCs that are applied to real and imaginary parts of the received signal, respectively.
Each ADC performs element-wise $B$-bit scalar quantization to the input signal. The quantization function of the scalar quantizer is denoted by $Q^\prime:\mathbb{R}\rightarrow\mathcal{Y}$, where $\mathcal{Y} \in \{q_1,q_2,\ldots,q_{2^B}\} $ is a set of quantization alphabets. For any real-valued input $r\in\mathbb{R}$, the quantization function outputs $Q^\prime(r)=q_k$ if $b_{k-1} \!<\! r \!\leq\! b_{k}$, where $b_k$ is the $k$-th quantization bin boundary such that $b_0\!=\!-\infty \!<\! b_1 \!<\! \ldots \!<\! b_{2^{B-1}} \!<\! b_{2^{B}}\!=\!\infty$. Using the above function, the received quantized vector after the ADCs at time slot $n$ is represented as 
\begin{align}\label{Eq_yn}
	{\bf y}[n] \!=\! Q({\bf r}[n]) \!=\! \left[\!\!\!\begin{array}{c}
		Q^\prime(r_{{\rm R},1}[n])\\ \vdots \\ Q^\prime(r_{{\rm R},2N_{\rm R}}[n])
	\end{array}\!\!\!\right]\!\!\in\! \mathcal{Y}^{2N_{\rm r}},
\end{align}
where $r_{{\rm R},i}[n]$ is the $i$th element of $[{\rm Re}({\bf r}[n])^{\!\top}\!, {\rm Im}({\bf r}[n])^{\!\top}]^{\!\top}$.

\section{Why Supervised Learning is Needed?}\label{sec:motivation}
In this section, we introduce the key concept of a supervised-learning-aided communication framework and then explain the motivation of this framework for a MIMO system with low-resolution ADCs.

\subsection{Concept}\label{sec:concept}
The key concept of the supervised-learning-aided framework is to learn the nonlinear input-output system formed by the concatenation of a wireless channel and a quantization function used at the ADCs, and then to use the learned information for data detection. We present this concept by using a simple example that consists of two phases: 1) system learning and 2) data detection.

\vspace{1mm}
\subsubsection{Considered Scenario}
In this example, we consider a real-coefficient MIMO channel with $N_{\rm t}=2$ and $N_{\rm r} = 2$, given by
\begin{align}
    {\bf H} = \left[
                \begin{array}{cc}
                  0.5 & 1 \\
                  1 & 0  \\
                \end{array}
              \right]
\end{align}
We assume that the receiver uses the one-bit ADCs with $\mathcal{Y}\!=\!\{-1,+1\}$. When BPSK modulation is used at the transmitter, i.e., $\mathcal{X}=\{-1,+1\}$,
the transmitter equipped with two transmit antennas is capable of sending four symbol vectors:
\begin{align}
    {\bf x}_1 \!=\! \left[\!\!
        \begin{array}{c}
            1 \\ 1 \\
        \end{array}\!\!\right]\!,~
    {\bf x}_2 \!=\! \left[\!\!\!
        \begin{array}{c}
            1 \\ -1 \\
        \end{array}\!\!\!\right]\!,~
    {\bf x}_3 \!=\! \left[\!\!\!
        \begin{array}{c}
            -1 \\ 1 \\
        \end{array}\!\!\!\right]\!,~\text{and}~
    {\bf x}_4 \!=\! \left[\!\!\!
        \begin{array}{c}
            -1 \\ -1 \\
        \end{array}\!\!\!\right]\!.
\end{align}
The set of the possible symbol vectors is denoted as $\mathcal{X}^2 = \left\{ {\bf x}_1, {\bf x}_2, {\bf x}_3, {\bf x}_4 \right\}$, and the index set of $\mathcal{X}^2$ is denoted as $\mathcal{K}=\{1,2,3,4\}$.

\vspace{1mm}
\subsubsection{System Learning Phase}
In the system learning phase, the transmitter sends all possible symbol vectors to the receiver by spanning $T_{\rm t} = 4$ time slots. In other words, the transmitter uses pilot signals defined as 
\begin{align}
	{\bf X}_{\rm t}= \big[ {\bf x}[1], {\bf x}[2], {\bf x}[3], {\bf x}[4]\big] =\big[{\bf x}_1, {\bf x}_2, {\bf x}_3, {\bf x}_4\big].
\end{align}
Under the premise that the noise signal is ignored during the learning  phase, the quantized vectors after the one-bit ADCs are received as follows:
\begin{align}
    {\bf y}[1] \!= \!\!\left[\!\!
        \begin{array}{c}
            1 \\ 1 \\
        \end{array}\!\!\right]\!\!,~
    {\bf y}[2] \!= \!\!\left[\!\!\!
        \begin{array}{c}
            -1 \\ 1 \\
        \end{array}\!\!\!\right]\!\!,~
    {\bf y}[3] \!= \!\!\left[\!\!\!
        \begin{array}{c}
            1 \\ -1 \\
        \end{array}\!\!\!\right]\!\!,~\text{and}~
    {\bf y}[4] \!= \!\!\left[\!\!\!
        \begin{array}{c}
            -1 \\ -1 \\
        \end{array}\!\!\!\right]\!.
\end{align}
By letting ${\bf y}_k \!=\! {\bf y}[k]$ for $k \!\in\! \mathcal{K}$, the set of the above four vectors is denoted by $\mathcal{Y}_{\rm t} \!=\! \left\{ {\bf y}_1, {\bf y}_2, {\bf y}_3, {\bf y}_4 \right\}$. By assuming that the receiver knows the transmitted pilot signals, the receiver is able to obtain a set $\mathcal{T} \!=\! \left\{({\bf y}_k,{\bf x}_k) | k\in\mathcal{K} \right\}$ of the four pairs of the quantized vector and the symbol vector. This set informs the input-output relations of a nonlinear system formed by a channel matrix ${\bf H}$ and the one-bit ADCs.

\vspace{1mm}
\subsubsection{Data Detection Phase}
By using $\mathcal{T}$, the receiver determines a mapping function $f: \mathcal{Y}^{2}\rightarrow \mathcal{K}$ that maps the received quantized vector to one of the indexes of possible symbol vectors. Then from the mapping function $f$, the receiver estimates which symbol vector was transmitted. One possible mapping-function design is to assign the index of the closest quantized vector in $\mathcal{Y}_{\rm t}$ to the received quantized vector at each time slot. This mapping function is represented as
\begin{align}\label{Eq III-A_ihat}
    f\left( {\bf y}[n]\right) = \argmin_{k\in \mathcal{K}}\|{\bf y}[n]-{\bf y}_k\|_2,
\end{align}
for $n \in \{5,6,\ldots,T\}$, where $\|\cdot\|_2$ is the Euclidean norm. From \eqref{Eq III-A_ihat}, the detected symbol vector is obtained as ${\hat {\bf x}}[n] = {\bf x}_{f\left( {\bf y}[n]\right)}$ for $ n \in \{5,6,\ldots, T\}$. For example, if the transmitter sends ${\bf x}_3= [-1,~1]^{\top}$ at time slot 5, the received quantized vector is ${\bf y}[5] = [1,-1]^{\top}$. From \eqref{Eq III-A_ihat}, the receiver chooses the index of the transmitted symbol vector as
\begin{align}
    f\left({\bf y}[5]\right)=\argmin_{k\in \mathcal{K}}\|{\bf y}[5]-{\bf y}_k\|_2\ = 3.
\end{align}
As a result, the receiver correctly estimates the transmitted symbol vector as $\hat{\bf x}[5] = {\bf x}_3$.

\vspace{2mm}
\noindent{\bf Remark 1} (Connection to supervised learning){\bf .}
	We can interpret the framework introduced in the above example through the lens of a classification problem in supervised learning. The determination of the mapping function $f$ using the training examples $({\bf y}_k,{\bf x}_k)$ for $k\!\in\!\mathcal{K}$ is equivalent to the design of a classifier $f: \mathcal{Y}_{\rm t} \!\rightarrow\! \mathcal{K}$ by using the training set $\mathcal{T} \!=\! \{({\bf y}_k,{\bf x}_k) | k \!\in\! \mathcal{K} \}$. Here, the index of the symbol vector and the quantized vector correspond to a class label and a feature vector, respectively. In this regard, the classifier $f$ serves as the detection rule, because the index of the transmitted symbol vector for the received quantized vector ${\bf y}[n]$, $n \!\ge\! 5$, is detected as $k^\star [n] \!=\! f({\bf y}[n])$. As a result, designing a good detection rule that accurately detects the symbol vector is equivalent to designing a good classifier that correctly assigns the class label.

\subsection{Motivation}\label{sec:SLmotiv}
The supervised-learning-aided communication framework is especially useful for solving a data detection problem in a MIMO system with low-resolution ADCs. In this system, conventional data detection methods, such as the optimal MLD \cite{Wang2014} and the GAMP-based algorithm \cite{Wen2016}, require both perfect CSIR and the knowledge of the quantization function used at the ADCs. Unfortunately, in practice, the accuracy of CSIR attained by a pilot-based channel estimation process is severely limited by a high quantization error at the ADCs \cite{Li2017,Studer2016,Wen2016}. In addition, the knowledge of the exact quantization function may not be available at the receiver when there exist hardware imperfections. Unlike the conventional methods, the proposed framework requires nor perfect CSIR or the knowledge of the quantization function, as it directly learns the nonlinear input-output relation between a transmitted symbol vector and a received quantized vector. Therefore, this framework has a potential to overcome the performance degradation of the conventional detection methods caused by inaccurate CSIR and/or the imperfect knowledge of the quantization function. 

\section{Supervised-Learning-Aided Communication Framework} \label{sec:SL}
In this section, by generalizing the concept introduced in Section \ref{sec:concept}, we propose the supervised-learning-aided communication framework for a MIMO system with low-resolution ADCs, which consists of two phases: 1) system learning and 2) data detection.

\subsection{System Learning Phase}\label{sec:training}
In the system learning phase, the receiver learns the input-output relations of a nonlinear system, formed by the concatenation of the wireless channel and the quantization function used at the ADCs, by utilizing $T_{\rm t}$ pilot signals. Unlike in the example in Section~\ref{sec:concept}, in practice, these input-output relations cannot be characterized by deterministic functions due to the existence of a random additive noise vector. To characterize the input-output relations including the randomness, we consider a conditional probability mass function (PMF) which is the probability of receiving a certain output for each possible candidate of input. Let ${\bf x}_k\!\in\!\mathcal{X}^{N_{\rm t}}$ be the $k$th possible symbol vector in a set $\mathcal{X}^{N_{\rm t}}$. Then the true conditional PMF of this nonlinear system for ${\bf x}_k$ is defined as
\begin{align}\label{Eq_IV-A_pdef}
	p({\bf y}|{\bf x}_k) = \mathbb{P}\big({\bf y}\!=\!Q({\bf H}{\bf x}_k\!+\!{\bf z})\big),
\end{align}
where ${\bf z}$ is a noise vector in which the elements are i.i.d. circularly-symmetric complex Gaussian random variables with zero mean and variance $\sigma^2$.

To learn the conditional PMF in \eqref{Eq_IV-A_pdef}, we develop two learning methods, referred to as \emph{full learning} and \emph{efficient learning} methods. The common idea of the developed methods is to send the repetitions of symbol vectors as pilot signals, so that the receiver observes multiple quantized vectors for every possible symbol vector. These multiple observations allow the receiver to empirically learn the conditional PMF for each symbol vector. Based on this idea, details of each method are described below.

\vspace{1mm}
\subsubsection{Full Learning Method}\label{sec:training1}
In the full learning method, the transmitter sends $L$ repetitions of all possible symbol vectors in $\mathcal{X}^{N_{\rm t}}$. A pilot-sequence matrix for this method is given by
\begin{align}\label{Eq IV-A_Xt}
	{\bf X}_{\rm t} &= \left[{\bf x}[1],~{\bf x}[2], \ldots, {\bf x}[T_{\rm t}] \right] \nonumber \\
	&= \big[\underbrace{{\bf x}_1,\ldots, {\bf x}_1}_{L~ {\rm repetitions}},{\bf x}_2,\ldots, {\bf x}_2,\ldots\ldots,{\bf x}_{K},\ldots, {\bf x}_{K}\big],
\end{align}
where $K\!=\!|\mathcal{X}|^{N_{\rm t}}\!=\!M^{N_{\rm t}}$.
After the pilot transmission, the receiver obtains $L$ quantized vectors for each possible symbol vector, under the premise that it has perfect knowledge of ${\bf X}_{\rm t}$. Motivated by the fact that the quantized vectors associating with the $k$th symbol vector are given by 
\begin{align}
	{\bf y}[(k\!-\!1)L\!+\!1], {\bf y}[(k\!-\!1)L\!+\!2], \ldots, {\bf y}[(k\!-\!1)L\!+\!L],
\end{align}
the receiver creates an empirical conditional PMF for the $k$th symbol vector as 
\begin{align}\label{Eq_IV-A_phat}
	{\hat p}({\bf y}| {\bf x}_k) \! =\! 
	\frac{1}{L} \sum_{t=1}^{L} {\bf 1}\!\left({\bf y} \!=\! {\bf y}[(k\!-\!1)L \!+\! t] \right),~\text{for}~{\bf y} \!\in\! \mathcal{Y}^{2N_{\rm r}},
\end{align}
where ${\bf 1}(A)$ is an indicator function that equals one if an event $A$ is true and zero otherwise. It is noticeable that 
the difference between ${\hat p}({\bf y}| {\bf x}_k)$ and $p({\bf y}|{\bf x}_k)$ vanishes as $L$ increases, by the law of large numbers.

For ease of exposition, we define the set of quantized vectors that are learned for the $k$th symbol vector as
\begin{align}\label{Eq_IV-A_Yhat}
	\mathcal{Y}_{{\rm t}, k} = \left\{ {\bf y} \big| {\hat p}({\bf y}|{\bf x}_k) \!>\! 0,~{\bf y}\!\in\!\mathcal{Y}^{2N_{\rm r}}\right\}.
\end{align}
We also define the total set of quantized vectors learned during this phase as $\mathcal{Y}_{\rm t} = \cup_{k=1}^{K}\mathcal{Y}_{{\rm t},k}  \subset \mathcal{Y}^{2N_{\rm r}}$.

\vspace{1mm} 
\subsubsection{Efficient Learning Method}\label{sec:training2}
As can be seen from \eqref{Eq IV-A_Xt}, the training overhead (i.e., the length of the pilot sequence) required for the full learning method is given by $T_{\rm t}=KL$. Somewhat surprisingly, this length can be reduced if 1) the system uses the QAM modulation and 2) the scalar quantization function is symmetric with respect to the origin, i.e., $Q^\prime(-r) \!=\! -Q^\prime(r)$  for $r\!\in\!\mathbb{R}$. To get an insight for this reduction, consider two symbol vectors ${\bf x}_{k_1}$ and ${\bf x}_{k_2}$ such that ${\bf x}_{k_1} \!=\!-{\bf x}_{k_2}$. Then the definition in \eqref{Eq_IV-A_pdef} implies that the conditional PMFs for ${\bf x}_{k_1}$ and ${\bf x}_{k_2}$ satisfy the following equality:
\begin{align}\label{Eq IV-A_p_eq}
	p({\bf y}|{\bf x}_{k_2})  
	&= \mathbb{P}\big({\bf y}\!=\!Q(-{\bf H}{\bf x}_{k_1} \!+\! {\bf z} )\big) \nonumber \\
	&= \mathbb{P}\big(-{\bf y}\!=\!Q({\bf H}{\bf x}_{k_1} \!-\! {\bf z} )\big)  \nonumber \\
	&= \mathbb{P}\big(-{\bf y}\!=\!Q({\bf H}{\bf x}_{k_1} \!+\! {\bf z} )\big)  \nonumber  \\
	&= p(-{\bf y}|{\bf x}_{k_1}).
\end{align}
Using a similar reasoning, we can also show that $p({\bf y}|{\bf x}_{k_3})\!=\! p({\bf y}^*|{\bf x}_{k_1})$ for ${\bf x}_{k_1}\!=\!j{\bf x}_{k_3}$, and $p({\bf y}|{\bf x}_{k_4})\!=\!p(-{\bf y}^*|{\bf x}_{k_1})$ for ${\bf x}_{k_1}\!=\!-j{\bf x}_{k_4}$, where 
\begin{align}\label{Eq IV-A_p_eq2}
	{\bf y}^* = 
	\left[\!\!\begin{array}{c}
		({\bf y})_{N_{\rm r}+1:2N_{\rm r}} \\ -({\bf y})_{1:N_{\rm r}}
	\end{array}\!\!\right],
\end{align}
and $({\bf y})_{a:b}$ with $a\leq b$ is a subvector of ${\bf y}$ that consists of its $a$th element to the $b$th element. These equalities allow the receiver to create the empirical conditional PMFs for three symbol vectors, $-{\mathbf{x}}_{k},j{\mathbf{x}}_{k},-j{\mathbf{x}}_{k}$, by utilizing the empirical conditional PMF for the $k$th symbol vector.

Motivated by the above observation, in the efficient learning method, the receiver first sets the indexes of the symbol vectors to satisfy that
\begin{align}\label{Eq IV-A_x_index}
	{\bf x}_k \!=\! -{\bf x}_{k+K/4} \!=\! j{\bf x}_{k+K/2} \!=\! -j{\bf x}_{k+3K/4},
\end{align}	 
for $k \!\in \!\left\{1,\ldots,{K}/{4}\right\}$. Based on this setting, a pilot-sequence matrix for the efficient method is given by
\begin{align}\label{Eq IV-A_Xt2}
	{\bf X}_{\rm t} &= \left[{\bf x}[1],~{\bf x}[2], \ldots, {\bf x}[T_{\rm t}] \right] \nonumber \\
	&= \Big[\underbrace{{\bf x}_1,\ldots, {\bf x}_1}_{L~ {\rm repetitions}},{\bf x}_2,\ldots, {\bf x}_2,\ldots\ldots,{\bf x}_{\frac{K}{4}},\ldots, {\bf x}_{\frac{K}{4}}\Big].
\end{align}
Then the receiver creates the empirical conditional PMF for the $k$th symbol vector as 
\begin{align}\label{Eq_IV-A_phat2}
	&{\hat p}({\bf y}| {\bf x}_k)  \nonumber \\
	&= \!\begin{cases}
		\frac{1}{L} \sum_{t=1}^{L}\! {\bf 1}\left( {\bf y} \!=\! {\bf y}[(k{-}1)L + t] \right), & \!\!\!\!k \!\in \!\left\{1,\ldots, \frac{K}{4}\right\}, \\
		{\hat p}(-{\bf y}| {\bf x}_{k-K/4}), & \!\!\!\! k \!\in \!\left\{\frac{K}{4}{+}1, \ldots, \frac{K}{2}\right\}, \\
		 {\hat p}({\bf y}^* |{\bf x}_{k-K/2}), & \!\!\!\! k \!\in \!\left\{\frac{K}{2}{+}1, \ldots, \frac{3K}{4}\right\}, \\
		{\hat p}(-{\bf y}^*| {\bf x}_{k-K/4}), & \!\!\!\! k \!\in \!\left\{\frac{3K}{4}{+}1, \ldots, {K}\right\}.\\
\end{cases}
\end{align}
As can be seen from \eqref{Eq IV-A_Xt2}, the training overhead of the efficient learning method is given by $T_{\rm t}\!=\!\frac{KL}{4}$, which is only a quarter of that of the full training method. Using this strategy, the receiver can effectively reduce the training overhead of the proposed framework when the system uses both the QAM modulation and the symmetric quantization function.

\subsection{Data Detection Phase} \label{sec:detection}
For the data detection phase, we develop two detection methods, referred to as \emph{eMLD} and \emph{MCD}. Both methods estimate transmitted symbol vectors by exploiting the empirical conditional PMFs obtained during the learning phase. Details of each method are described below.

\vspace{1mm}
\subsubsection{Empirical-Maximum-Likelihood Detection (eMLD)}
The key idea of eMLD is to select the index of a symbol vector that maximizes the empirical conditional PMF, namely,
\begin{align}\label{Eq_eMLD0}
    k^\star[n] = \argmax_{k} ~{\hat p}({\bf y}[n]| {\bf x}_k).
\end{align}
When the number of training repetitions in the learning phase approaches infinity (i.e., $L\rightarrow \infty$),  the empirical distribution converges to the corresponding true distribution by the law of large numbers. In this ideal case, by the definition, the eMLD method in \eqref{Eq_eMLD0} is the optimal MLD method of the MIMO system with low-resolution ADCs, regardless of the number of precision bits at the ADCs.

\begin{figure}[t]
	\centering
	\subfigure[Empirical-maximum-likelihood detection (eMLD)]
	{\epsfig{file=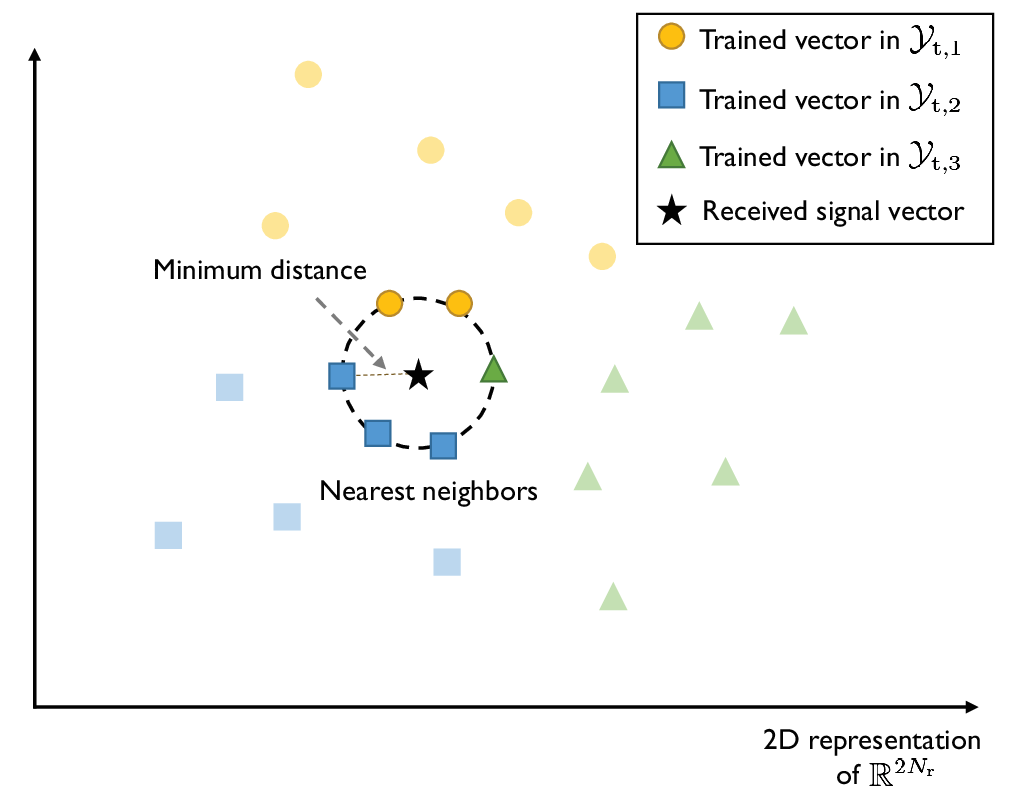, width=6.1cm}}
	\subfigure[Minimum-center-distance detection (MCD)]
	{\epsfig{file=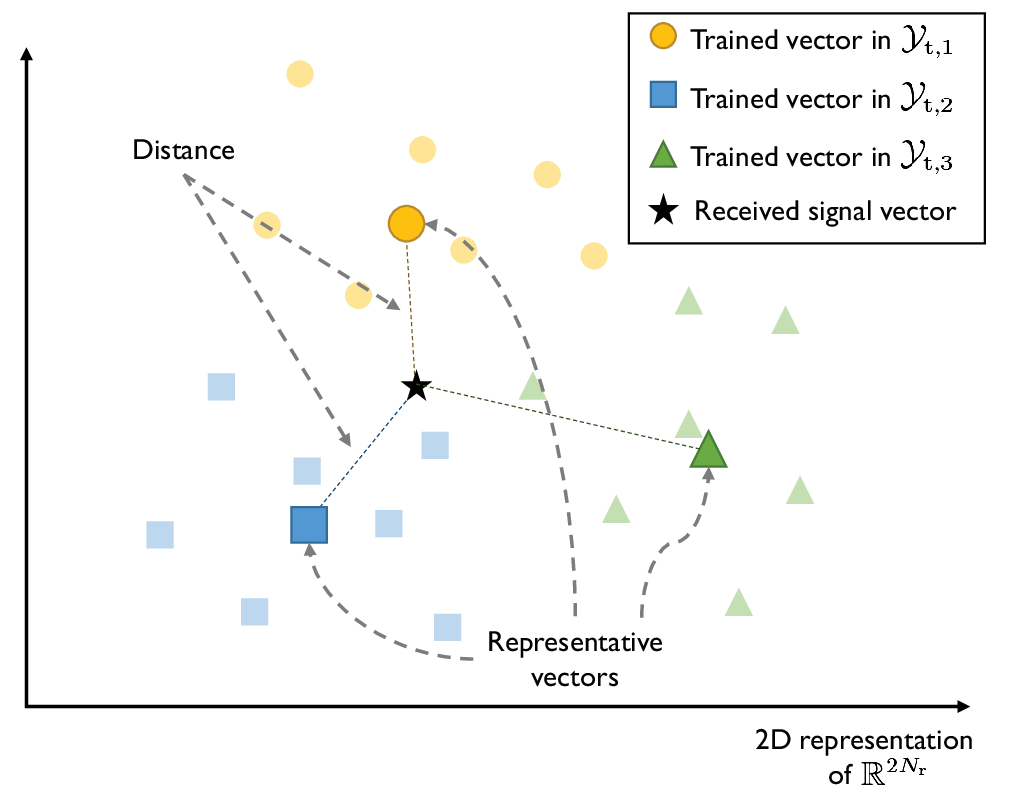, width=6.1cm}}
	\caption{Illustration for key concepts of the developed data detection methods (eMLD and MCD) when $K=3$.}
	\label{fig:concept}
\end{figure}

Unfortunately, when $L$ is insufficient, eMLD is no more optimal for two reasons: 1) the empirical conditional PMF is different from the true conditional PMF, and 2) there is a non-zero probability event that a received quantized vector during the detection phase has not been learned during the previous learning phase, which obviously results in a detection failure. To resolve these problems, we extend the detection rule in \eqref{Eq_eMLD0} by considering a set of quantized vectors in $\mathcal{Y}_{{\rm t}}$ that are closest to the received quantized vector. Let $\mathcal{N}({\bf y}[n])$ be the set of the closest vectors to ${\bf y}[n]$ among the vectors in $\mathcal{Y}_{{\rm t}}$ with respect to the Euclidean distance, i.e.,
\begin{align}\label{Eq_neighbor}
    \mathcal{N}({\bf y}[n]) = \left\{ {\bf y}_{\rm t}~ \Big| \left\|{\bf y}[n] - {\bf y}_{\rm t}\right\|_2 =  R_{\rm min}[n],~{\bf y}_{\rm t} \in \mathcal{Y}_{{\rm t}}  \right\},
\end{align}
where $R_{\rm min}[n] = \min_{{\bf y}_{\rm t} \in \mathcal{Y}_{\rm t}} \left\|{\bf y}[n] - {\bf y}_{\rm t}\right\|_2$. Using this set, the detection rule of the eMLD method, $f_{\rm eMLD}:\mathcal{Y}^{2N_{\rm r}} \rightarrow \mathcal{K}=\{1,2,\ldots,K\}$, is given by
\begin{align}\label{Eq_eMLD}
      f_{\rm eMLD}({\bf y}[n]) = \argmax_{k} \sum_{{\bf y} \in \mathcal{N}({\bf y}[n])} \hat{p}({\bf y} |{\bf x}_k).
\end{align}
The eMLD method in \eqref{Eq_eMLD} is illustrated in Fig. \ref{fig:concept}(a). Note that when $L$ is sufficiently large, the detection rule in  \eqref{Eq_eMLD} becomes equivalent to \eqref{Eq_eMLD0}.

\vspace{2mm}
\noindent{\bf Remark 2} (Connection to a $K$-nearest neighbors classifier){\bf .}
	The eMLD method resembles with a $K$-nearest-neighbors (KNN) classifier which is widely used to solve the classification problem in supervised learning. The key idea of the KNN classifier is that when an unlabeled vector is observed, it finds the $K$-nearest neighbors to the observed vector, and assigns a label to the vector by using the majority voting of neighbors' labels. As explained, eMLD finds the neighbor set $\mathcal{N}({\bf y}[n])$, then assigns the index $f_{\rm eMLD}({\bf y}[n])$ as the most probably index for the vectors in $\mathcal{N}({\bf y}[n])$. Therefore, the eMLD method is similar to the KNN classifier in the sense that they simply compare the number of neighbors' labels. One notable difference is that eMLD uses the neighbor set of elements that are equidistant from the received vector.

\vspace{1mm}
\subsubsection{Minimum-Center-Distance Detection (MCD)}
Although eMLD is optimal for the case of $L\!=\!\infty$, its computational complexity may not be acceptable for use in practical systems when the size of $\mathcal{Y}_{{\rm t}}$ is large. The reason is that the receiver requires to compute all distances among the received quantized vector and the vectors in $\mathcal{Y}_{{\rm t}}$. To resolve this problem, we present a simple detection method, called MCD, that requires a less detection complexity than the eMLD method.

The key idea of MCD is to create a set of $K$ representative vectors at the receiver for the detection as depicted in Fig. \ref{fig:concept}(b). 
The receiver creates a representative quantized vector for the $k$th symbol vector by computing the conditional expectation with respect to the empirical PMF, i.e.,
\begin{align}\label{Eq_barY}
    \bar{\bf y}_{{\rm t},k} &\triangleq {\mathbb{E}}_{{\bf y}_{\rm t}}[{\bf y}_{\rm t}|{\bf x}={\bf x}_k] = \sum_{{\bf y}_{\rm t} \in \mathcal{Y}_{{\rm t},k}} {\bf y}_{\rm t}\hat{p}({\bf y}_{\rm t}|{\bf x}_k) \in \mathbb{R}^{2N_{\rm r}}.
\end{align}
Notice that the representative vectors are not necessarily an element of $\mathcal{Y}^{2N_{\rm r}}$.
Utilizing $K$ representative vectors, the MCD method, $f_{\rm MCD}: \mathcal{Y}^{2N_{\rm r}}\rightarrow \mathcal{K}$, finds the index that minimizes the distance between ${\bf y}[n]$ and  $\bar{\bf y}_{{\rm t},k}$ as follows:
\begin{align}\label{Eq_MCD}
    f_{\rm MCD}({\bf y}[n]) = \argmin_{k}\left\| {\bf y}[n] - \bar{\bf y}_{{\rm t},k} \right\|_2.  
\end{align}

\vspace{2mm}
\noindent{\bf Remark 3} (Connection to a nearest-centroid classifier){\bf .}
	The principle of MCD is very close to that of a nearest-centroid classifier (NCC) which is a simple solution of the classification problem in supervised learning. NCC assigns the class label of a unlabeled observed vector by using the centroid vectors that represent their classes. Similarly, MCD determines the index of the detected symbol vector as the index with the minimum distance from the conditional mean vector of the quantized vectors that are already learned, each of which is associated with an input symbol vector. This resemblance is a good example to show an interesting connection between a data detection problem in wireless communications and a classification problem in supervised learning.

\section{Supervised-Learning-Aided Successive-Interference-Cancellation}\label{sec:SIC}
One drawback of the supervised-learning-aided framework in Section~\ref{sec:SL} is that it is not affordable in practical communication systems when the number of transmit antennas $N_{\rm t}$ or the modulation size $M$ is large. The reason is that both the training overhead and the computational complexity of this framework exponentially increase with $M$ and $N_{\rm t}$. To overcome this drawback, in this section, we develop a supervised-learning-aided successive-interference-cancellation (SL-SIC) for data detection in a MIMO system with low-resolution ADCs. The developed SL-SIC reduces both the training overhead and the computational complexity of the proposed framework in Section~\ref{sec:SL}.

The key idea of SL-SIC is to divide a symbol vector into two subvectors with reduced dimensions, and then to detect these two subvectors successively using the proposed framework with MCD. Based on this idea, the SL-SIC method consists of three phases: 1) symbol vector division, 2) system learning, and 3) data detection. Details of each phase are described below.

\subsection{Symbol Vector Division Phase}
In the symbol vector division phase, the receiver divides the symbol vector ${\bf x}[n]$ into two subvectors ${\bf x}^{(1)}[n] \in \mathcal{X}^{N_{\rm t,1}}$ and ${\bf x}^{(2)}[n] \in \mathcal{X}^{N_{\rm t,2}}$, where $N_{\rm t,1} \in \mathbb{N}$ and $N_{\rm t,2} \in \mathbb{N}$ such that $N_{\rm t} = N_{\rm t,1} + N_{\rm t,2}$. Then the received quantized vector after the ADCs at time slot $n$ is rewritten using these two subvectors as
\begin{align}\label{Eq_y_divide}
    {\bf y}[n] = Q\left({\bf H}^{(1)}{\bf x}^{(1)}[n] + {\bf H}^{(2)}{\bf x}^{(2)}[n] + {\bf z}[n] \right),
\end{align}
where ${\bf H}^{(1)} \in \mathbb{C}^{N_{\rm r}\times N_{\rm t,1}}$ and ${\bf H}^{(2)} \in \mathbb{C}^{N_{\rm r}\times N_{\rm t,2}}$ are the channel sub-matrices associated with the first subvector and the second subvector, respectively.

Our strategy for the symbol vector division is to maximize the chordal distance between the subspace spanned by ${\bf H}^{(1)}$ and that spanned by ${\bf H}^{(2)}$ in \eqref{Eq_y_divide}. The purpose of this strategy is to minimize the effect of the second subvector on the detection of the first subvector. To realize this strategy, we adopt a pilot-based channel estimation process, unlike the supervised-learning-aided communication framework presented in Section~\ref{sec:SL}. We then develop an algorithm that determines the column indexes of the channel matrix associating with each subvector, based on the estimated channel matrix. Let  $\hat{\bf H}$ and $\hat{\bf h}_{k}$ be the estimated channel matrix and its $k$th column vector. Also, let ${\bf U}_{\mathcal{I}} \!\in\! \mathbb{C}^{N_{\rm r}\times |\mathcal{I}|}$ be the orthonormal basis matrix of the subspace spanned by the columns ${\bf \hat h}_{\mathcal{I}(1)},\ldots, {\bf \hat h}_{\mathcal{I}(|\mathcal{I}|)}$ for $\mathcal{I} \subset \{1,\ldots,N_{\rm t}\}$. Using these notations, the developed algorithm is summarized as in Algorithm~\ref{alg:DetermineI}.

\begin{algorithm}{\small{
	\caption{The proposed algorithm for symbol vector division}
	\label{alg:DetermineI}
	\begin{algorithmic}[1]
		\STATE Set $\mathcal{I}^{(1)}=\emptyset$ and $\mathcal{I}^{(2)}=\{1,2,\ldots,N_{\rm t}\}$.
		\FOR {$i  = 1$ \TO $N_{\rm t,1}$}
		\STATE Find $k^{\star}\!=\!\argmax_{k\in \mathcal{I}^{(2)}} d_{\rm chordal}\left({\bf U}_{\mathcal{I}^{(1)}\cup\{k\}}, {\bf U}_{\mathcal{I}^{(2)}\setminus\{k\}}\right)$, 
				where $d_{\rm chordal}\left({\bf A},{\bf B}\right) = \frac{1}{\sqrt{2}} \|{\bf A}{\bf A}^H \!-\! {\bf B}{\bf B}^H\|_{\rm F}$.
		\STATE Update $\mathcal{I}^{(1)} \leftarrow \mathcal{I}^{(1)} \cup \{k^{\star}\}$ and $\mathcal{I}^{\rm (2)} \leftarrow \mathcal{I}^{\rm (2)} \setminus \{k^{\star}\}$.
		\ENDFOR
	\end{algorithmic}}}
\end{algorithm}
In Step 3, the proposed algorithm selects the column index in $\mathcal{I}^{(2)}$ that maximizes the chordal distance between two subspaces spanned by the column vectors corresponding to $\mathcal{I}^{(1)}\cup \{k\}$ and $\mathcal{I}^{(2)}\setminus\{k\}$, respectively. Then in Step 4, the proposed algorithm adds the selected index to the set $\mathcal{I}^{(1)}$, while discarding the selected index from the set $\mathcal{I}^{(2)}$.  Steps~3$\sim$4 are repeated for $N_{\rm t,1}$ times. From Algorithm~\ref{alg:DetermineI}, the receiver obtains the two sub-matrices, each associated with the $u$th subvector as follows:
\begin{align}
    \hat{\bf H}^{(u)} = \left[\hat{\bf h}_{i_1^{(u)}},\hat{\bf h}_{i_2^{(u)}},\cdots, \hat{\bf h}_{i_{N_{\rm t,1}}^{(u)}} \right],
\end{align}
where $i_j^{(u)}$ is the $j$th element of $\mathcal{I}^{(u)}$ for $u \in \{1,2\}$.

\subsection{System Learning Phase}
In the system learning phase, the receiver empirically learns a conditional PMF of the received quantized vector for each possible first subvector. The key difference to the learning phase in Section~\ref{sec:training} is that the conditional PMF is now \emph{marginalized} for all possible second subvectors, since the receiver does not have the information of the transmitted second subvector at the time of the learning phase. In addition to this difference, we also consider the \emph{effective} received vector that is projected onto the orthogonal subspace spanned by ${\bf H}^{(2)}$ after the ADCs, in order to suppress the effect of the second symbol vector on the marginalized conditional PMF. Based on this strategy, we first define the effective received vector\footnote{Unlike in MIMO systems with infinite-resolution ADCs, the effective received vector in \eqref{Eq_projected_y} is corrupted by the interference signals of ${\bf H}^{(2)}{\bf x}^{(2)}$ even after the orthogonal projection, because of 1) the nonlinearity of the quantization function $Q(\cdot)$ and 2) the imperfect channel estimation, i.e., $\hat{\bf H}^{(2)} \neq {\bf H}^{(2)}$. Nevertheless, the orthogonal-projection approach can still be used to \textit{suppress} the effect of the interference caused by the second subvector.} at time slot $n$ as
\begin{align}\label{Eq_projected_y}
	{\bf \tilde y}[n] = {\bf W} Q\left({\bf H}^{(1)}{\bf x}^{(1)}[n] + {\bf H}^{(2)}{\bf x}^{(2)}[n] + {\bf z}[n] \right),
\end{align}
where ${\bf W} = \big(\hat{\bf H}_{\rm R}^{(2)}\big)^\perp\in \mathbb{R}^{(2N_{{\rm r}}-2N_{\rm t,2})\times 2N_{\rm r}}$  is a projection matrix whose rows are the orthogonal basis of the left null space of
\begin{align}\label{Eq_realeqR}
	\hat{\bf H}_{\rm R}^{(2)} = \left[ \begin{array}{cc}
	{\rm Re}(\hat{\bf H}^{(2)}) \!\!\!\!& -{\rm Im}(\hat{\bf H}^{(2)}) \\
	{\rm Im}(\hat{\bf H}^{(2)}) \!\!\!\!& {\rm Re}(\hat{\bf H}^{(2)}) \\
	\end{array} \right].
\end{align}
We then define the marginalized conditional PMF of the effective received vector for the $k$th possible candidate of the first subvector as 
\begin{align}\label{Eq_marginPMF}
	&{p}({\bf \tilde y}| {\bf x}_{k}^{(1)}) = \sum_{j=1}^{K_2} {p}\big({\bf \tilde  y}\big| {\bf x}_{k}^{(1)},{\bf x}_{j}^{(2)}\big)
	{p}\big({\bf x}_{j}^{(2)}\big) \nonumber \\
	&= \frac{1}{K_2}\sum_{j=1}^{K_2} \mathbb{P}\Big({\bf \tilde  y}\!=\!{\bf W} Q\big({\bf H}^{(1)}{\bf x}_k^{(1)} + {\bf H}^{(2)}{\bf x}_j^{(2)} + {\bf z}\big)\Big), 
\end{align}
for $k \!\in\! \{1,\ldots,K_1\!=\!M^{N_{{\rm t},1}}\}$ and ${\bf \tilde y} \!\in\! \mathbb{R}^{2N_{\rm r}-2N_{\rm t,2}}$, where ${\bf x}_k^{(1)} \!\in\! \mathcal{X}^{N_{\rm t,1}}$ is the $k$th possible first subvector, ${\bf x}_{j}^{(2)} \!\in\! \mathcal{X}^{N_{\rm t,2}}$ is the $j$th possible second subvector, and $K_2 = M^{N_{{\rm t},2}}$. 
Unfortunately, learning the pair-wise conditional PMF in \eqref{Eq_marginPMF} using the methods in Section~\ref{sec:training} still entails high training overhead when $K_1K_2$ is large. In addition, the pilot signals of the system are already utilized by a channel estimation process during the symbol vector division phase. Therefore, for the SL-SIC method, we develop an alternative learning method, called \textit{pseudo} learning, that does not require additional training signals beyond the pilot signals utilized for the channel estimation.

The key idea of pseudo learning is to artificially generate multiple received vectors based on the estimated channel, instead of actually sending the training signals for it. These generated vectors are used to create the empirical conditional PMFs by taking the role of received signals in the original learning methods. For this, the receiver generates $L$ artificial effective received vectors for $({\bf x}_{k}^{(1)},{\bf x}_{j}^{(2)})$, in which the $\ell$th artificial vector is given by
\begin{align}\label{Eq_artificial_y_1st}
    &{\bf \hat y}_{k,j}^{(1,\ell)} =
    {\bf W} Q\left(\hat{\bf H}^{(1)}{\bf x}_{k}^{(1)} + \hat{\bf H}^{(2)}{\bf x}_{j}^{(2)} + {\bf \hat z}^{(\ell)}\right),
\end{align}
where ${\bf \hat z}^{(\ell)} \in \mathbb{C}^{N_{\rm r}}$ is the $l$th artificial noise vector whose elements are independently generated from  $\mathcal{CN}(0,{\sigma^2})$ for $\ell \in \{1,\ldots,L\}$. By using these artificial vectors, the empirical conditional PMF for ${\bf x}_{k}^{(1)}$ marginalized with respect to all possible second subvectors is created as
\begin{align}\label{Eq_marginPMF_1st}
    {\hat p}\big({\bf \tilde y}\big| {\bf x}_{k}^{(1)}\big) 
    = \frac{1}{K_2 L} \sum_{j=1}^{K_2} \sum_{\ell=1}^{L} {\bf 1}\left( {\bf \tilde y}= {\bf \hat y}_{k,j}^{(1,\ell)}\right),
\end{align}
for ${\bf \tilde y}\!\in\! \mathbb{R}^{2N_{\rm r}-2N_{\rm t,2}}$. Then the set of effective received vectors learned for ${\bf x}_{k}^{(1)}$ is given by
\begin{align}\label{Eq_Yhat_art}
	\mathcal{Y}_{{\rm t}, k}^{(1)} = \left\{ {\bf \tilde y} \Big| {\hat p}\big({\bf \tilde y}\big|{\bf x}_{k}^{(1)}\big) \!>\! 0,~{\bf \tilde y}\!\in\! \mathbb{R}^{2N_{\rm r}-2N_{\rm t,2}}\right\}.
\end{align}
As shown in the above, the SL-SIC method requires both CSIR and the knowledge of the quantization function used at the ADCs, unlike the supervised-learning-aided framework in Section \ref{sec:SL}. Nevertheless, it still has some advantages over conventional data detection methods (e.g., the optimal MLD method \cite{Wang2014} or the GAMP-based algorithm \cite{Wen2016}) in terms of the detection performance or the computational complexity, which will be explained in the sequel.

\subsection{Data Detection Phase}
For the data detection phase, we develop a two-stage MCD method that consists of two successive MCD methods. I the developed method, the receiver estimates the first subvector based on the empirical conditional PMFs obtained during the learning phase, then estimates the second subvector by using the estimated first subvector. Details of two-stage MCD are described below.

The receiver detects the first symbol subvector by applying the MCD method with the empirical conditional PMFs obtained during the system learning phase. As in the original MCD method, the receiver creates  a representative received vector for each possible first subvector. The representative received vector for the $k$th possible first subvector, namely $\bar{\bf y}_{{\rm t},k}^{(1)} \in \mathbb{R}^{2N_{\rm r}-2N_{\rm t,2}}$, is obtained as
\begin{align}\label{Eq_barY_1st}
    \bar{\bf y}_{{\rm t},k}^{(1)}
    = \sum_{{\bf y}_{\rm t} \in \mathcal{Y}_{{\rm t},k}^{(1)}} {\bf y}_{\rm t}\hat{p}({\bf y}_{\rm t}|{\bf x}_{k}^{(1)}).
\end{align}
Using the representative vectors, the detection rule for the first subvector, $f_{\rm MCD,1}: \mathcal{Y}^{2N_{\rm r}}\rightarrow \{1,\ldots,K_1\}$, is given by 
\begin{align}\label{Eq_TMCD1}
    f_{\rm MCD,1}({\bf y}[n]) = \argmin_{k} \big\| {\bf W}{\bf y}[n] - \bar{\bf y}_{{\rm t},k}^{(1)} \big\|_2,
\end{align}
From \eqref{Eq_TMCD1}, the estimated first subvector at time slot $n$ is obtained as $\hat{\bf x}^{(1)}[n] = {\bf x}_{k^\star[n]}^{(1)}$, where $k^\star[n] = f_{\rm MCD,1}({\bf y}[n])$.

After detecting the first subvector, the receiver again applies the MCD method to detect the second symbol subvector. For this, the receiver first adopts the pseudo learning method to learn the conditional PMFs for all possible second subvectors, under the assumption that the estimated first subvector is transmitted. Then the empirical conditional PMF for ${\bf x}_{j}^{(2)}$ is created as 
\begin{align}\label{Eq_marginPMF_2nd}
	{\hat p}\big({\bf y}\big| {\bf x}_{j}^{(2)}\big) 
	= \frac{1}{L} \sum_{\ell=1}^{L} {\bf 1}\left( {\bf y}= {\bf \hat y}_{j}^{(2,\ell)}\right),
\end{align}\noindent
for ${\bf y}\!\in\! \mathcal{Y}^{2N_{\rm r}}$, where the $l$th artificial vector is
\begin{align}\label{Eq_artificial_y_2nd}
	&{\bf \hat y}_{j}^{(2,\ell)} =
	Q\left(\hat{\bf H}^{(1)}\hat{\bf x}^{(1)}[n] + \hat{\bf H}^{(2)}{\bf x}_{j}^{(2)} + {\bf \hat z}^{(\ell)}\right),
\end{align}
as in \eqref{Eq_artificial_y_1st}. Using the learned conditional PMFs, the detection rule for the second subvector, $f_{\rm MCD,2}: \mathcal{Y}^{2N_{\rm r}}\rightarrow \{1,\ldots,K_2\}$, is represented as
\begin{align}\label{Eq_TMCD2}
    f_{\rm MCD,2}({\bf y}[n]) = \argmin_{j} \big\| {\bf y}[n] -\bar{\bf y}_{{\rm t},j}^{(2)} \big\|_2,
\end{align}
where $\bar{\bf y}_{{\rm t},j}^{(2)}$ is the representative received vector for the $j$th possible second subvector
\begin{align}\label{Eq_barY_2nd}
	\bar{\bf y}_{{\rm t},j}^{(2)}
	= \sum_{{\bf y}_{\rm t} \in \mathcal{Y}_{{\rm t},j}^{(2)}} {\bf y}_{\rm t}\hat{p}({\bf y}_{\rm t}|{\bf x}_{j}^{(2)}),
\end{align}
and $\mathcal{Y}_{{\rm t},j}^{(2)}\!=\! \{ {\bf y} | {\hat p}({\bf y}|{\bf x}_{j}^{(2)})\!>\!0,{\bf y}\!\in\!\mathcal{Y}^{2N_{\rm r}}\}.$ From \eqref{Eq_TMCD2}, the estimated second subvector at time slot $n$ is obtained as $\hat{\bf x}^{(2)}[n] = {\bf x}_{k^\star[n]}^{(2)}$, where $k^\star[n] = f_{\rm MCD,2}({\bf y}[n])$.

The symbol vector transmitted at time slot $n$, ${\bf x}[n]$, can be reconstructed from two estimated symbol subvectors, $\hat{\bf x}^{(1)}[n]$ and $\hat{\bf x}^{(2)}[n]$, for $n \in \{T_{\rm t}+1,\ldots,T\}$. Let $\hat{\bf x}[n] =[\hat{x}_{1}[n],\hat{x}_{2}[n],\cdots,\hat{x}_{N_{\rm t}}[n]]^\top$ be the estimated symbol vector at time slot $n$. For an index $m \in \{1,2,\ldots,N_{\rm t}\}$, if $m$ corresponds to the $j$th element of $\mathcal{I}^{(u)}$, the $m$th element of $\hat{\bf x}[n]$ is determined as
    \begin{align}\label{Eq_x_TMCD}
        \hat{x}_{m}[n]  = \hat{x}_{j}^{(u)}[n]   ~~\text{for}~~n\!\in\! \{T_{\rm t}\!+\!1,\ldots,T\},
    \end{align}
where $\hat{x}_{j}^{(u)}[n]$ is the $j$th element of $\hat{\bf x}^{(u)}[n]$ for $u \in \{1,2\}$.

The detection complexity of the developed SL-SIC method has the order of $\mathcal{O}(M^{\max\{N_{{\rm t},1},N_{{\rm t},2}\}})$. This complexity order is less than those of the proposed framework with MCD or the optimal MLD, which are given by $\mathcal{O}(M^{N_{\rm tx}})$.

\section{Analysis for MIMO systems With One-Bit ADCs} \label{sec:analysis}
In this section, we characterize the detection performance of the supervised-learning-aided communication framework presented in Section~\ref{sec:SL} for a MIMO system with one-bit ADCs.

\subsection{Upper Bound of Vector-Error-Rate}\label{sec:analysis:VER}
This section characterizes an upper bound of  VER for the proposed framework with the MCD method. In particular,  the upper bound is derived under an ideal assumption referred to as \textit{perfect learning} which implies that the receiver perfectly learns the received quantized vectors for all possible symbol vectors, namely,
\begin{align}
	\bar{\bf y}_{{\rm t},k} = Q({\bf H}{\bf x}_k),~~\text{for}~~k\in \mathcal{K}. \label{Eq_perf_train}
\end{align}
Under this assumption, the following theorem provides the upper bound of VER.
\vspace{+2mm}
\begin{thm}\label{Thm1}
    Suppose MIMO systems with one-bit ADCs. Under the perfect learning assumption in \eqref{Eq_perf_train}, the upper bound of the VER for the proposed framework with MCD is
    \begin{align}\label{Eq_Thm1}
        P_{\rm e}^{\rm vec}
        \le \!\frac{1}{M^{N_{\rm t}}}\!\sum_{k=1}^{M^{N_{\rm t}}}\sum_{j=D_k}^{2N_{\rm r}} &\!\!\sum_{i=1}^{\binom{2N_{\rm r}}{j}}
        \!\prod_{l\in \mathcal{S}_{i,j}}\!\! \left(1-\Phi\!\left(\sqrt{\frac{2\rho|g_{k,l}|^2}{N_{\rm t}}}\right)\right)\nonumber \\
        &~~~~~~\times\!\!\prod_{l^\prime \notin \mathcal{S}_{i,j}} \!\!\Phi\!\left(\sqrt{\frac{2\rho|g_{k,l^\prime}|^2}{N_{\rm t}}}\right)\!,
    \end{align}
    where $\rho = \frac{N_{\rm t}}{\sigma^2}$ is the SNR of the system, $\mathcal{S}_{i,j}$ is the $i$th possible subset of $\{1,2,\ldots,2N_{\rm r}\}$ with size $j$,
        $D_k = \min_{i\neq k} \left\lfloor \frac{d_{k,i}+1}{2} \right\rfloor$, 
        $d_{k,i} = \left\| Q({\bf H}{\bf x}_k ) -  Q({\bf H}{\bf x}_i ) \right\|_0$, and
        $g_{k,l}$ is the $l$th element of ${\bf g}_k = \big[{\rm Re}\left( {\bf H}{\bf x}_k  \right)^{\top}\!\!,~ {\rm Im}\left( {\bf H}{\bf x}_k  \right)^{\top} \big]^{\top}$.
\end{thm}
\begin{IEEEproof}
    In this proof, we omit the index $n$ of time slot for ease of exposition.
    Suppose that the receiver equipped with one-bit ADCs adopts the MCD method.
    Then the receiver detects the symbol vector as ${\bf \hat x}={\bf x}_{k^\star}$,
    where ${k^\star}=f_{\rm MCD}({\bf y})$, and ${\bf y}=Q({\bf H}{\bf x}+{\bf z})\in \{-1,+1\}^{2N_{\rm r}}$.
    Let $P_{{\rm e},k}^{\rm vec} = \mathbb{P}\left( \hat{\bf x} \neq {\bf x}_{k} | {\bf x} = {\bf x}_k  \right)$ be the pair-wise error probability that the detected symbol vector is different from ${\bf x}_{k}$ when the transmitter sends ${\bf x}_{k}$. Then VER is defined as
        \begin{align}\label{Eq_V-A_Perr}
            P_{\rm e}^{\rm vec} = \sum_{k=1}^{M^{N_{\rm t}}} \mathbb{P}\left( \hat{\bf x} \neq {\bf x}_{k}, {\bf x} = {\bf x}_k \right)  = \frac{1}{M^{N_{\rm t}}} \sum_{k=1}^{M^{N_{\rm t}}} P_{{\rm e},k}^{\rm vec}.
        \end{align}
	With the  perfect learning assumption in \eqref{Eq_perf_train} and the use of one-bit ADCs, the detection rule of MCD in \eqref{Eq_MCD} is rewritten as
    \begin{align}
       k^\star &= \argmin_{k}  \left\| {\bf y}   - Q({\bf H}{\bf x}_k)  \right\|_2   \\
        &= \argmin_{k} \left\| {\bf y}   - Q({\bf H}{\bf x}_k)   \right\|_0 \label{Eq_V-A_MCD},
    \end{align}
    where $ \| {\bf a}\|_0$ is the zero norm that denotes the number of nonzero elements in a vector ${\bf a}$. Note that the equality of \eqref{Eq_V-A_MCD} holds only for the one-bit-ADC case.
    From \eqref{Eq_V-A_MCD}, $P_{{\rm e},k}^{\rm vec}$ of the MCD method is upper bounded as
    \begin{align}\label{Eq_V-A_Pei}
        P_{{\rm e},k}^{\rm vec} &\le \mathbb{P}\Big(\! \| {\bf y} {-} Q({\bf H}{\bf x}_k)  \|_0  \ge \min_{j\neq k}\| {\bf y} {-} Q({\bf H}{\bf x}_j)  \|_0 \Big| {\bf x} = {\bf x}_k  \!\Big).
    \end{align}
    For further analysis, we define a set $\mathcal{C}=\{ Q({\bf H}{\bf x}_1), Q({\bf H}{\bf x}_2),\ldots, Q({\bf H}{\bf x}_K) \}$ which is interpreted as an error-correcting code where each element $Q({\bf H}{\bf x}_k)$ can be treated as a codeword vector of $\mathcal{C}$. For any code, one can define the distance between two codes $Q({\bf H}{\bf x}_k)$ and $Q({\bf H}{\bf x}_i)$ as $d_{k,i}= \left\| Q({\bf H}{\bf x}_k ) -  Q({\bf H}{\bf x}_i ) \right\|_0$. Then $\left\| {\bf y} - Q({\bf H}{\bf x}_k) \right\|_0 \ge \big\lfloor \frac{d_{k,i}+1}{2} \big\rfloor$ is the necessary condition for an event that the MCD method outputs ${\bf x}_i$ when ${\bf x}_k$ was sent. Thus, we obtain an upper bound\footnote{Although this upper bound is loose in general, it is useful to reveal the key features of $P_{{\rm e},k}^{\rm vec}$.} as
    \begin{align} \label{Eq_V-A_Pei2}
        &\!\!P_{{\rm e},k}^{\rm vec}
        \le \mathbb{P}\!\left(\! \bigcup_{i=1, i\neq k}^{K}\!\! \left\{\left\| {\bf y} - Q({\bf H}{\bf x}_k) \right\|_0 \ge\! \left\lfloor \frac{d_{k,i}{+}1}{2} \right\rfloor\right\}
            \bigg| {\bf x} = {\bf x}_k \!\right)  \nonumber \\
        &\!\!= \mathbb{P}\left( \left\| {\bf y} - Q({\bf H}{\bf x}_k) \right\|_0 \ge D_{k} \big| {\bf x} = {\bf x}_k \right)  \nonumber \\
        &\!\!= \mathbb{P} \left( \left\| Q({\bf H}{\bf x}_k + {\bf z})- Q({\bf H}{\bf x}_k) \right\|_0 \ge D_{k} \right) \nonumber \\
        &\!\!= \mathbb{P}\left( \sum_{l=1}^{2N_{\rm r}} {\bf 1}\left( {\rm sign}(g_{k,l} +z_{{\rm R},l}) \neq {\rm sign}(g_{k,l}) \right) \ge D_{k} \right),
    \end{align}
    where $D_k = \min_{i\neq k} \left\lfloor \frac{d_{k,i}{+}1}{2} \right\rfloor$,
        ${\rm sign}(\cdot)$ is the signum function,
        $g_{k,l}$ is the $l$th element of ${\bf g}_k$,
        $z_{{\rm R},l}$ is the $l$th element of ${\bf z}_{\rm R}$,
    \begin{align}\label{Eq_V-A_gi}
        {\bf g}_k
        = \left[
            \begin{array}{c}
                {\rm Re}\left( {\bf H}{\bf x}_k \right) \\
                {\rm Im}\left( {\bf H}{\bf x}_k \right) \\
            \end{array}
        \right]\!,~~\text{and}~~
        {\bf z}_{\rm R}
        = \left[
            \begin{array}{c}
                {\rm Re}( {\bf z} ) \\
                {\rm Im}( {\bf z} ) \\
            \end{array}
        \right]\!.
    \end{align}
    Because $z_{{\rm R},l}$ is i.i.d. as $\mathcal{N}(0,\frac{\sigma^2}{2})$ for all $l$, the probability of an event that the sign of $g_{k,l}$ is flipped due to the noise $z_{{\rm R},l}$ is given by
    \begin{align}\label{Eq_V-A_p_pair}
    	p_{k,l}^{\rm e} =1-\Phi\left(\sqrt{\frac{2\rho|g_{k,l}|^2}{N_{\rm t}}}\right).
    \end{align}	
    Using this fact, \eqref{Eq_V-A_Pei2} is rewritten as
    \begin{align}\label{Eq_V-A_upper2}
        &\mathbb{P}\left( \sum_{l=1}^{2N_{\rm r}} {\bf 1}\left( {\rm sign}(g_{k,l} +z_{{\rm R},l}) \neq {\rm sign}(g_{k,l}) \right)  \ge  D_k \right)
        \nonumber \\
        &= \sum_{j=D_k}^{2N_{\rm r}} \sum_{i=1}^{\binom{2N_{\rm r}}{j}}\prod_{l\in \mathcal{S}_{i,j}} p_{k,l}^{\rm e}
            \prod_{l^\prime \notin \mathcal{S}_{i,j}} (1-p_{k,l^\prime}^{\rm e}).
    \end{align}
    Plugging  \eqref{Eq_V-A_Pei2} and \eqref{Eq_V-A_upper2} to \eqref{Eq_V-A_Perr} yields \eqref{Eq_Thm1}. This completes the proof.
\end{IEEEproof}

\vspace{2mm}
The upper bound of VER in \eqref{Eq_Thm1} can be interpreted as the effective error probability of an error correcting code $\mathcal{C}=\{ Q({\bf H}{\bf x}_1), Q({\bf H}{\bf x}_2),\ldots, Q({\bf H}{\bf x}_K) \}$, in which the $l$th layer of the $k$th codeword has the error probability of $p_{k,l}^{\rm e}$. With this interpretation, we further simplify the upper bound in \eqref{Eq_Thm1}, to provide more clear understanding on the VER of MCD, especially in high SNR regime. The result is given in the following Corollary:

\begin{cor}\label{Cor1}
	Suppose MIMO systems with one-bit ADCs. Under the perfect learning assumption in \eqref{Eq_perf_train}, the logarithm of VER is upper bounded as
	\begin{align}\label{Eq_Cor1}
		\ln ({P}_{\rm e}^{\rm vec}) \le -\frac{D_{\rm min} |g_{\rm min}|^2}{N_{\rm t}}\rho + O(1),
	\end{align}
	where $D_{\rm min} = \min_k D_k$, $g_{\rm min} = \min_{(k,l)} g_{k,l}$, and $O(1)$ is an expression that does not depend on SNR.
\end{cor}
\begin{IEEEproof}
	Because $p_{k,l}^{\rm e}$ in \eqref{Eq_V-A_p_pair} is the decreasing function of $|g_{k,l}|$, the right-hand side (RHS) of \eqref{Eq_Thm1} is further upper bounded as
	\begin{align}
		P_{\rm e}^{\rm vec}
		&\le \frac{1}{M^{N_{\rm t}}} \sum_{k=1}^{M^{N_{\rm t}}}  \sum_{j=D_k}^{2N_{\rm r}} \binom{2N_{\rm r}}{j}  \prod_{l=l_{k,1}^{\star}}^{l_{k,j}^{\star}}  p_{k,l}^{\rm e}  \label{Eq_upper2} \\
		&\le C_{N_r,D_{\rm min}} \Bigg\{1-\Phi\Bigg(\sqrt{\frac{2\rho |g_{\rm min}|^2}{N_{\rm t}}}\Bigg)\Bigg\}^{\!\!D_{\rm min}} \\
		&\le C_{N_r,D_{\rm min}} \exp{\left(-\frac{D_{\rm min} |g_{\rm min}|^2}{N_{\rm t}} \rho \right)},\label{Eq_upper3}
	\end{align}
	where $D_{\rm min} = \min_k D_k$,
	$C_{N_r,D_{\rm min}} = \sum_{j=D_{\rm min}}^{2N_{\rm r}} \binom{2N_{\rm r}}{j}$,
	$l_{k,i}^{\star}$ is the index of the element of ${\bf g}_{k}$ that has the $i$th-minimum absolute value,
	and $g_{\rm min} = \min_{(k,l)} g_{k,l}$. By taking the logarithm to \eqref{Eq_upper3}, we arrive at the result in \eqref{Eq_Cor1}, where $O(1) = \ln (C_{N_r,D_{\rm min}})$.
\end{IEEEproof}

\vspace{2mm}
Corollary~\ref{Cor1} demonstrates that the upper bound of VER decreases exponentially with SNR $\rho$, the minimum channel gain $|g_{\rm min}|^2$, the inverse of $N_{\rm t}$, and the half of the minimum distance, $D_{\rm min}=\left\lfloor \frac{d_{\rm min}+1}{2} \right\rfloor$. The most interesting parameter here is $d_{\rm min}$, which represents how far the transmitted symbol vectors are separated in a received domain. For a certain channel realization ${\bf H}$, the received signals from two different symbol vectors ${\bf x}_i$ and ${\bf x}_j$ can be identical even without noise, i.e., $Q({\bf H}{\bf x}_i )=Q({\bf H}{\bf x}_j)$ for $i\neq j$. In this case, these two vectors cannot perfectly be distinguished by any detection method. The upper bound in \eqref{Eq_Cor1} also agrees with this fact because the RHS of \eqref{Eq_Cor1} becomes a constant when $d_{\rm min}=0$.

To reduce VER, it is important to design the communication system to have a large enough minimum distance $d_{\rm min}$. One simple way is to increase the number of receive antennas. For example, if $N_{\rm t}=2$, we have four possible symbol vectors $\{{\bf x}_1,{\bf x}_2,{\bf x}_3,{\bf x}_4\}\in\{-1,+1\}^2$ that generate a code $\mathcal{C}=\{ Q({\bf H}{\bf x}_1), Q({\bf H}{\bf x}_2),Q({\bf H}{\bf x}_3), Q({\bf H}{\bf x}_4) \}\in \{-1,+1\}^{2N_{\rm r}}$. Clearly, the minimum distance $d_{\rm min}$ of $\mathcal{C}$ increases with $N_{\rm r}$, because each codeword can be mapped into a higher-dimensional space. This characteristic can be interpreted as a \textit{receive diversity gain} in the MIMO system with one-bit ADCs. The relation between $d_{\rm min}$ and $N_{\rm r}$ will be more clearly shown in the sequel.

\vspace{2mm}
\noindent{\bf Remark 4} (Realization of perfect learning assumption){\bf .}
	We explain how to realize the perfect learning assumption in \eqref{Eq_perf_train} using the full learning method introduced in Section~\ref{sec:training}. In this method, the representative vector for the $k$th symbol vector is obtained as 
	\begin{align}\label{Eq_V-A_bary}
	\bar{\bf y}_{{\rm t},k} =\frac{1}{L}\sum_{\ell=1}^{L} Q({\bf H}{\bf x}_k + {\bf z}).
	\end{align}
	Specifically, the $l$th element of $\bar{\bf y}_{{\rm t},k}$ in \eqref{Eq_V-A_bary} is given by 
	\begin{align}\label{Eq_V-A_baryl}
	\bar{y}_{{\rm t},k,l} =\frac{1}{L}\sum_{\ell=1}^{L} {\rm sign}(g_{k,l} + {z}_{{\rm R},l} ).
	\end{align}    
	where ${z}_{{\rm R},l}$ is the $l$th element of ${\bf z}_{\rm R} = \big[{\rm Re}\left( {\bf  z} \right)^{\top}\!\!,~ {\rm Im}\left( {\bf z} \right)^{\top} \big]^{\top}$. Because the empirical probability converges to the corresponding true probability by the law of large numbers, as the number of training repetitions goes to infinity (i.e., $L\rightarrow \infty$),
	\begin{align}\label{Eq_V-A_bary_converge}
	\bar{y}_{{\rm t},k,l}  
	&\rightarrow  \mathbb{P}\left(  {\rm sign}(g_{k,l} + {z}_{{\rm R},l} ) = {\rm sign}(g_{k,l}) \right) {\rm sign}(g_{k,l})\nonumber \\
	&\qquad +  \mathbb{P}\left(  {\rm sign}(g_{k,l} + {z}_{{\rm R},l} ) \neq {\rm sign}(g_{k,l}) \right) (-{\rm sign}(g_{k,l}))  \nonumber \\
	&= \left\{2\Phi\left(\sqrt{\frac{2\rho|g_{k,l}|^2}{N_{\rm t}}}\right)-1 \right\} {\rm sign}(g_{k,l}),
	\end{align} 
	where the first inequality is obtained from \eqref{Eq_V-A_p_pair}. Therefore, for sufficiently large SNR such that $\Phi\Big(\!\sqrt{\frac{2\rho|g_{k,l}|^2}{N_{\rm t}}}\Big) \approx 1$ for all $k$ and $j$, the full learning method obtains the result in \eqref{Eq_perf_train}.

\subsection{Distribution of $d_{\rm min}$ for Rayleigh-Fading Channel}\label{sec:analysis:dmin}
We have shown that the detection error probability of the proposed framework depends on the minimum distance, $d_{\rm min}$, which is closely related to a channel realization. To provide a clear understanding for this minimum distance under a random channel realization, we derive the distribution of $d_{\rm min}$ by assuming Rayleigh-fading channels and BPSK modulation.

\begin{thm}\label{Thm2}
    Suppose MIMO systems with one-bit ADCs and BPSK modulation. For Rayleigh-fading channels, the complementary cumulative distribution function (CCDF) of $d_{\rm min}$ is approximated as
    \begin{align}\label{Eq_Thm2}
        \mathbb{P}\left( d_{\rm min} \ge n \right)
        \!\approx \!\!\prod_{1\le i < j \le 2^{N_{\rm t}-1}}\!\!\! \sum_{k=n}^{2N_{\rm r}-n}\!\!\!
            \binom{2N_{\rm r}}{k}\!\!  \left(1{-}p_{{\rm eq},\delta_{i,j}}\right)^k \!p_{{\rm eq},\delta_{i,j}}^{2N_{\rm r}-k},
    \end{align}
    where $\delta_{i,j}=\|{\bf x}_i - {\bf x}_j\|_0$, and $p_{{\rm eq},\delta} = \frac{2}{\pi}{\rm arctan}\left(\sqrt{\frac{N_{\rm t} - \delta}{\delta}}\right)$, under the premise that the indexes of possible symbol vectors satisfy ${\bf x}_{2^{N_{\rm t}}-k+1} = -{\bf x}_k$ for $k \in \{1,\ldots, 2^{N_{\rm t}-1} \}$. The approximation in \eqref{Eq_Thm2} becomes an equality if $N_{\rm t}=2$ and becomes less accurate as $N_{\rm t}$ increases.
\end{thm}
\begin{IEEEproof}
    When the BPSK modulation is used for the transmission, for any possible symbol vector ${\bf x} \in \{+1,-1\}^{N_{\rm t}} $, we have $-{\bf x} \in \{+1,-1\}^{N_{\rm t}}$. From this fact, we can set the indexes of the symbol vectors to satisfy that ${\bf x}_{2^{N_{\rm t}}-k+1} = -{\bf x}_k$ for $k \in \{1,\ldots, 2^{N_{\rm t}-1} \} $.  Then by the definition of $d_{\rm min}$, the CCDF of $d_{\rm min}$ is represented as 
    
    \vspace{-5mm}{\small{\begin{align}\label{Eq_V-B_Pdmin}
        &\mathbb{P}\left( d_{\rm min} \ge n \right)
            = \mathbb{P}\left(\! \bigcap_{1\le i< j \le 2^{N_{\rm t}}}\!\! \left\{\| Q({\bf H}{\bf x}_i) - Q({\bf H}{\bf x}_j) \|_0 \ge n \right\} \!\right) \nonumber \\
        &= \mathbb{P}\Bigg( \bigcap_{1\le i < j \le 2^{N_{\rm t}{-}1}}
        \bigg[ \min\Big\{ \| Q({\bf H}{\bf x}_i) - Q({\bf H}{\bf x}_j) \|_0,
        \nonumber\\
        &\qquad \qquad \qquad\qquad\qquad\| Q({\bf H}{\bf x}_i) - Q(-{\bf H}{\bf x}_j) \|_0 \Big\} \ge n \bigg] \Bigg).
    \end{align}}}\noindent
    When the receiver is equipped with the one-bit ADCs defined with $\mathcal{Y}\!=\!\{-1,+1\}$, the following equality holds:
    \begin{align}\label{Eq_V-B_Qsum}
        \| Q({\bf H}{\bf x}_i) \!-\! Q({\bf H}{\bf x}_j) \|_0 \!+\! \| Q({\bf H}{\bf x}_i) \!+\! Q({\bf H}{\bf x}_j) \|_0 \!=\! 2N_{\rm r}.
    \end{align}
    Applying \eqref{Eq_V-B_Qsum} to \eqref{Eq_V-B_Pdmin} with $Q(-{\bf H}{\bf x}_j) = -Q({\bf H}{\bf x}_j)$ yields
    
    \vspace{-3mm}{\small{\begin{align}\label{Eq_V-B_Pdmin2}
        &\mathbb{P}\left( d_{\rm min} \ge n \right)
            \nonumber\\
        &= \mathbb{P}\vast( \!\bigcap_{1\le i < j \le 2^{N_{\rm t}-1}}\!\!
        \Big\{ \underbrace{n \le \| Q({\bf H}{\bf x}_i) {-} Q({\bf H}{\bf x}_j) \|_0 \le 2N_{\rm r} {-} n}_{\triangleq E_{i,j}} \Big\}\! \vast).
    \end{align}}}\noindent
    Unfortunately, further simplification of the RHS of \eqref{Eq_V-B_Pdmin2} is very difficult due to the complicated dependence of events $E_{i,j}$ for different $i$ and $j$.
    Therefore, in this work, we only provide the approximation of \eqref{Eq_V-B_Pdmin2} by ignoring the statistical dependence among the events $\{E_{i,j}\}_{i,j}$:
    
    \vspace{-3mm}{\small{\begin{align}\label{Eq_V-B_Papprox}
        &\mathbb{P}\left( d_{\rm min} \ge n \right) 
        =\mathbb{P}\left( \!\bigcap_{1\le i < j \le 2^{N_{\rm t}-1}}\! E_{i,j} \right) \approx \prod_{1\le i < j \le 2^{N_{\rm t}-1}}\mathbb{P}(E_{i,j}) .
    \end{align}}}\noindent
    The approximation in \eqref{Eq_V-B_Papprox} becomes an equality if $N_{\rm t}=2$ and becomes less accurate as $N_{\rm t}$ increases.
    The probability of each pair event $E_{i,j}$ in \eqref{Eq_V-B_Papprox} is calculated as
    \begin{align}\label{Eq_V-B_Eij0}
        &\mathbb{P}(E_{i,j}) 
        =\mathbb{P} \left( n \le \| Q({\bf H}{\bf x}_i) - Q({\bf H}{\bf x}_j) \|_0 \le 2N_{\rm r} - n   \right)  \nonumber\\
        &= \sum_{k=n}^{2N_{\rm r}-n} \mathbb{P}\left(  \| Q({\bf H}{\bf x}_i) - Q({\bf H}{\bf x}_j) \|_0 = k \right) \nonumber \\
        &= \sum_{k=n}^{2N_{\rm r}-n} \mathbb{P}\left( \sum_{l=1}^{2N{\rm r}}  {\bf 1}\left( {\rm sign}(g_{i,l}) \neq {\rm sign}(g_{j,l})\right) = k \right).
    \end{align}
    For Rayleigh-fading channels, each channel element is drawn from an i.i.d. circularly-symmetric complex Gaussian random variable with zero mean and unit variance. Therefore, for each symbol vector ${\bf x}_i \in \{-1,+1\}^{N_{\rm t}}$, all elements of ${\bf g}_i$ are i.i.d. as $\mathcal{N}\left(0,\frac{N_{\rm t}}{2}\right)$.
    Using this fact, the RHS of \eqref{Eq_V-B_Eij0} is rewritten as
    \begin{align}\label{Eq_V-B_Eij}
        &\mathbb{P}(E_{i,j}) = \sum_{k=n}^{2N_{\rm r}-n} \binom{2N_{\rm r}}{k} \prod_{l=1}^{k} \mathbb{P}\left( {\rm sign}(g_{i,l}) \neq {\rm sign}(g_{j,l}) \right)  \nonumber \\
            &\qquad\qquad \qquad~~~\times\prod_{l=k+1}^{2N_{\rm r}} \mathbb{P}\left( {\rm sign}(g_{i,l}) = {\rm sign}(g_{j,l}) \right).
    \end{align}
    Let $u_{i,j,l} = \frac{g_{i,l} + g_{j,l}}{2}$, $v_{i,j,l} = \frac{g_{i,l} - g_{j,l}}{2}$, and $\delta_{i,j}=\|{\bf x}_i - {\bf x}_j\|_0$ be the number of different elements between two symbol vectors ${\bf x}_i $ and ${\bf x}_j$.
    Then the distributions of $u_{i,j,l}$ and $v_{i,j,l}$ are given by
    \begin{align}\label{Eq_V-B_uvdist}
        u_{i,j,l}  \sim \mathcal{N}\left(0,\frac{N_{\rm t}-\delta_{i,j}}{2}\right),
        ~\text{and}~ v_{i,j,l}  \sim \mathcal{N}\left(0,\frac{\delta_{i,j}}{2}\right),
    \end{align}
    for all $l \in \{1,\ldots,2N_{\rm r}\}$.
    From \eqref{Eq_V-B_uvdist} and the definitions of $u_{i,j,l}$ and $v_{i,j,l}$, we obtain
    \begin{align}\label{Eq_V-B_sign}
        &\mathbb{P}\left( {\rm sign}(g_{i,l}) = {\rm sign}(g_{j,l}) \right)  \nonumber \\
        &= \mathbb{P}\left( {\rm sign}(u_{i,j,l} + v_{i,j,l}) = {\rm sign}(u_{i,j,l} - v_{i,j,l}) \right) \nonumber \\
        &= \int_{-\infty}^{\infty} \frac{1}{\sqrt{\pi(N_{\rm t}-\delta_{i,j})}} {e}^{-\frac{u^2}{N_{\rm_t}-\delta_{i,j}}}
            \int_{-|u|}^{|u|}  \frac{1}{\sqrt{\pi\delta_{i,j}}} {e}^{-\frac{v^2}{\delta_{i,j}}} dv du \nonumber \\
        &=\frac{2}{\pi}{\rm arctan}\left(\sqrt{\frac{N_{\rm t} - \delta_{i,j}}{\delta_{i,j}}}\right)  = p_{{\rm eq},\delta_{i,j}}.
    \end{align}
    Substituting \eqref{Eq_V-B_sign} into \eqref{Eq_V-B_Eij} and then applying the result to \eqref{Eq_V-B_Papprox} yields \eqref{Eq_Thm2}.
    This completes the proof.
\end{IEEEproof}

\vspace{2mm}
Although Theorem~\ref{Thm2} only provides the approximate CCDF of $d_{\rm min}$, this result is still useful to understand how $d_{\rm min}$ behaves with the number of receive and transmit antennas. In particular, Theorem~\ref{Thm2} provides the exact CCDF of $d_{\rm min}$ when $N_{\rm t}=2$. Therefore, in this special case, we clearly reveal that how $d_{\rm min}$ changes with $N_{\rm r}$ by using the following corollary:

\begin{cor}\label{Cor2}
   When $N_{\rm t}=2$, the probability that $d_{\rm min}$ is larger than $n=cN_{\rm r}$ asymptotically goes to one, for any $0 \le c < 1$, i.e.,
    \begin{align}\label{Eq_Cor2}
        \lim_{N_{\rm r} \rightarrow \infty}\mathbb{P}\left( d_{\rm min} \ge c N_{\rm r}\right) = 1.
    \end{align}
\end{cor}
\begin{IEEEproof}
	When $N_{\rm t} = 2$, the approximation in \eqref{Eq_Thm2} becomes an equality. By using this fact, \eqref{Eq_Cor2} is rewritten as 
	    \begin{align}\label{Eq_V-B_Pdmin_2Nt}
	        &\mathbb{P}\left( d_{\rm min} \ge n \right) \nonumber \\
	        &= \sum_{k=n}^{2N_{\rm r}-n} \binom{2N_{\rm r}}{k} \!\left(\frac{2}{\pi}{\rm arctan}(1)\right)^{\!2N_{\rm r}-k}\!
	            \left(1-\frac{2}{\pi}{\rm arctan}(1)\right)^k \nonumber \\
	        &= \sum_{k=n}^{2N_{\rm r}-n}\binom{2N_{\rm r}}{k} 2^{-2N_{\rm r}},
	    \end{align}
	    where the first equality is obtained from $\delta_{1,2}=1$. 
	    
    Theorem 5.3.2 in \cite{LovaszBook} says that for $0 \le t \le m$,

    \vspace{-3mm}{\small{\begin{align}\label{Eq_V-B_Lovasz}
        \sum_{k=0}^{m-t-1} \!\binom{2m}{k} 2^{-2m} \!+\!\! \sum_{k=m+t+1}^{2m}\! \binom{2m}{k} 2^{-2m}  \le \exp\left(-\frac{t^2}{m+t}\right),
    \end{align}}}where $m$ is a positive integer.
    Because $\sum_{k=0}^{2N_{\rm r}}\binom{2N_{\rm r}}{k} 2^{-2N_{\rm r}}=1$, applying the inequality in \eqref{Eq_V-B_Lovasz} to \eqref{Eq_Cor2} yields
    \begin{align}\label{Eq_V-B_Pdmin_LB}
        \mathbb{P}\left( d_{\rm min} \ge n \right) \ge  1 -  \exp\left(-\frac{(N_{\rm r}-n)^2}{2N_{\rm r}-n}\right).
    \end{align}
    Let $n=cN_{\rm r}$ for any $0 \le c \le 1$. Then the lower bound of \eqref{Eq_V-B_Pdmin_LB} becomes
    \begin{align}\label{Eq_V-B_Pdmin_LB2}
        \mathbb{P}\left( d_{\rm min} \ge n \right) \ge  1 -  \exp\left(-\frac{(1-c)^2}{2-c}N_{\rm r}\right).
    \end{align}
    Except for $c=1$, the RHS of the above inequality goes to one as $N_{\rm r}$ increases, so we obtain the results in \eqref{Eq_Cor2}.
\end{IEEEproof}

\vspace{2mm}
As shown in Corollary~\ref{Cor2}, the probability of an event that $d_{\rm min}$ is larger than an arbitrarily close value of $N_{\rm r}$ goes to one as $N_{\rm r}$ increases. This result implies that $d_{\rm min}$ is an increasing function of $N_{\rm r}$ for a sufficiently large value of $N_{\rm r}$, which also agrees with our intuition. Therefore, by combining the results in Corollaries 1 and 2, we are able to show that the VER of the proposed framework decreases as the number of receive antennas increases for the MIMO system with one-bit ADCs.

\section{Numerical Results} \label{sec:simul}
In this section, using simulations, we evaluate the detection performance of the supervised-learning-aided framework proposed in Section~\ref{sec:SL} and also the SL-SIC method developed in Section~\ref{sec:SIC} for a MIMO system with low-resolution ADCs. We also validate the analysis results in Section~\ref{sec:analysis} by simulations.

\vspace{-1mm}
\subsection{Performance Evaluation for the Proposed Methods} 
For the performance evaluation, we consider the symbol-error-rate (SER) performance achieved by the proposed and existing detection methods. We assume Rayleigh-fading channels, so each element of the channel matrix is independently drawn from $\mathcal{CN}(0,1)$. We design the scalar quantizer of the ADCs to maximize the output entropy as in \cite{Messerschmitt:71}. For this, we assume that each real and imaginary part of an input signal is a Gaussian random variable with zero-mean and variance of $\sigma_x^2\!=\!\frac{N_{\rm t}+\sigma^2}{2}$. Under this assumption, we determine the bin boundaries of the quantizer to satisfy that $b_0\!=\!-\infty \!<\! b_1 \!<\! \ldots \!<\! b_{2^{B-1}} \!<\! b_{2^{B}}\!=\!\infty$  with
	\begin{align}
	\Phi\left( \frac{b_k}{\sigma_x}\right)-\Phi\left( \frac{b_{k-1}}{\sigma_x}\right) &= \frac{1}{2^B}.
	\end{align}
We also determine the $k$th quantization alphabet as 
\begin{align}
	q_k &= \frac{\int_{u=b_{k-1}}^{u=b_k} \frac{u}{\sqrt{2\pi\sigma_x^2}} e^{-{\frac{u^2}{2\sigma_x^2}}} du}
		{\int_{u=b_{k-1}}^{u=b_k} \frac{1}{\sqrt{2\pi\sigma_x^2}} e^{-{\frac{u^2}{2\sigma_x^2}}} du}.
\end{align}

\begin{figure*}[t]
	\centering
	\subfigure[$T_{\rm t}\!=\!12$ ($L\!=\!3$)]
	{\epsfig{file=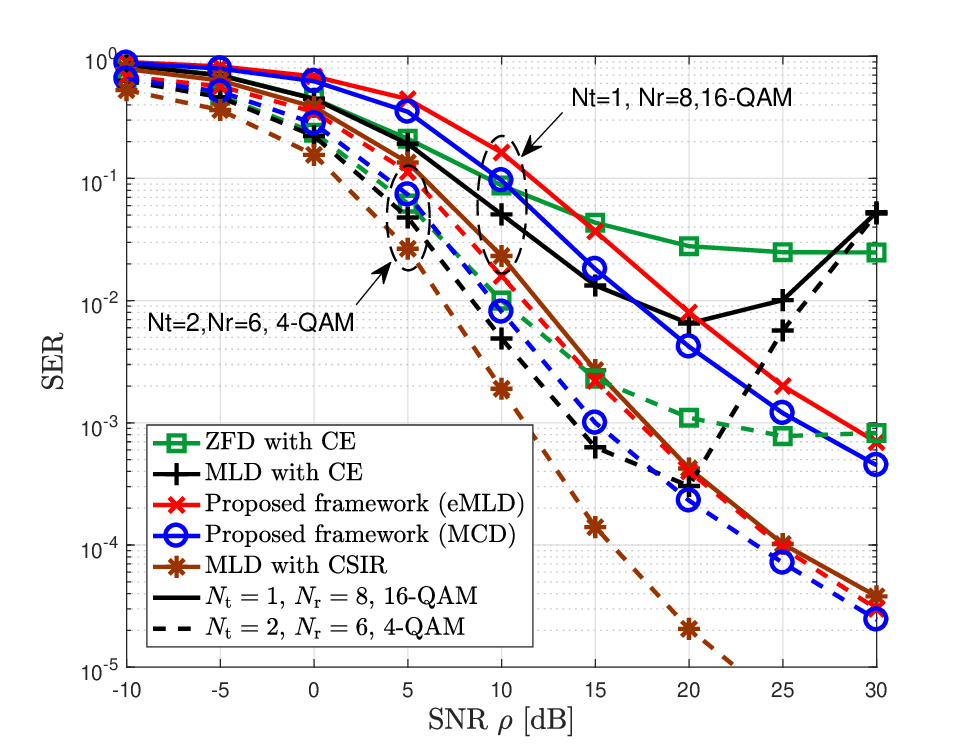, width=6.8cm}}
	\subfigure[$T_{\rm t}\!=\!48$ ($L\!=\!12$)]
	{\epsfig{file=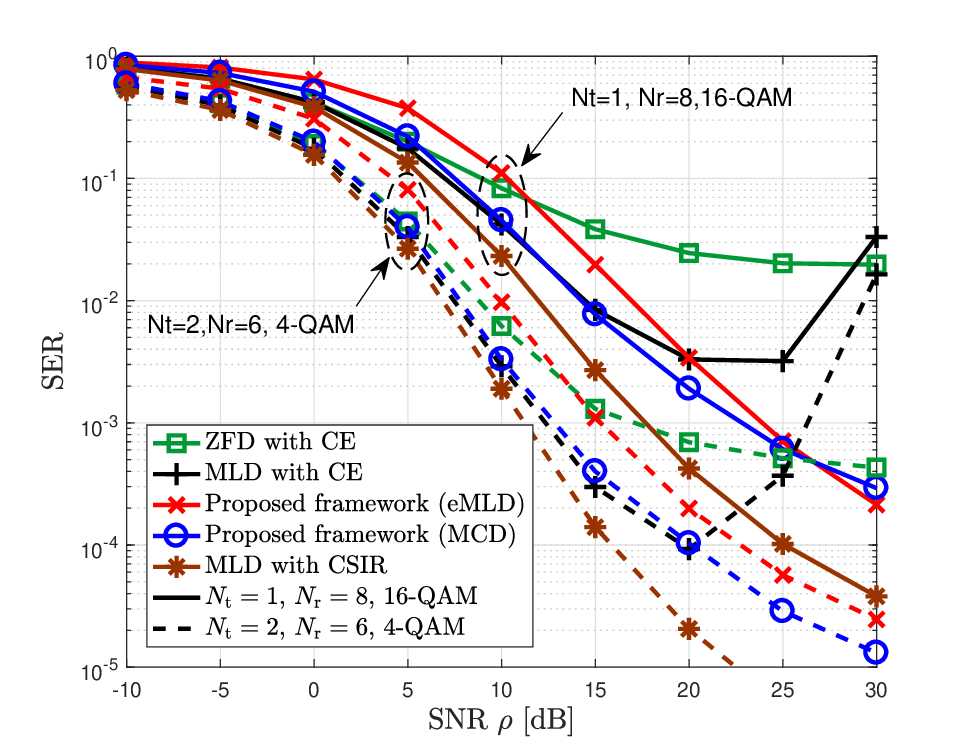, width=6.8cm}}
	\caption{SER vs. SNR of the proposed framework and conventional detection methods in a MIMO system with 2-bit ADC ($B\!=\!2$). For the conventional detection methods with channel estimation (CE), the least-squares-based CE method in \cite{Risi2016} is adopted with pilot signals of length $T_{\rm t}$. For the proposed framework, the efficient learning method in Section~\ref{sec:training} is adopted with $L\!=\!4T_{\rm t}/K$ training repetitions.}\vspace{-3mm}
	\label{fig:SL}
\end{figure*}
\begin{figure*}[t]
	\centering
	\subfigure[2-bit ADC ($B\!=\!2$)]
	{\epsfig{file=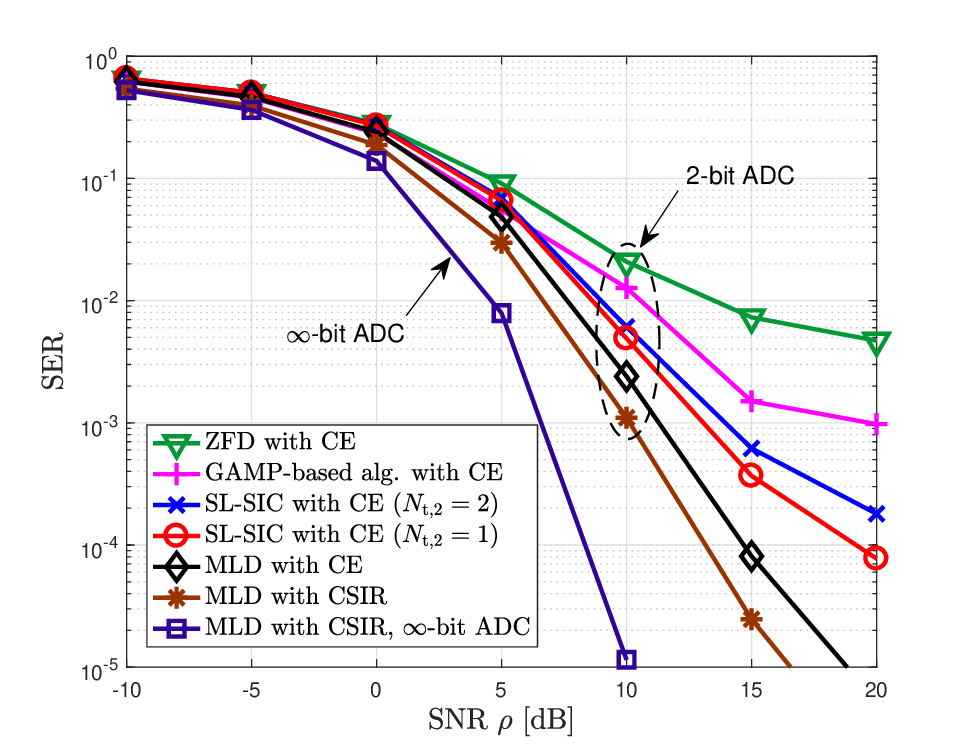, width=6.8cm}}
	\subfigure[3-bit ADC ($B\!=\!3$)]
	{\epsfig{file=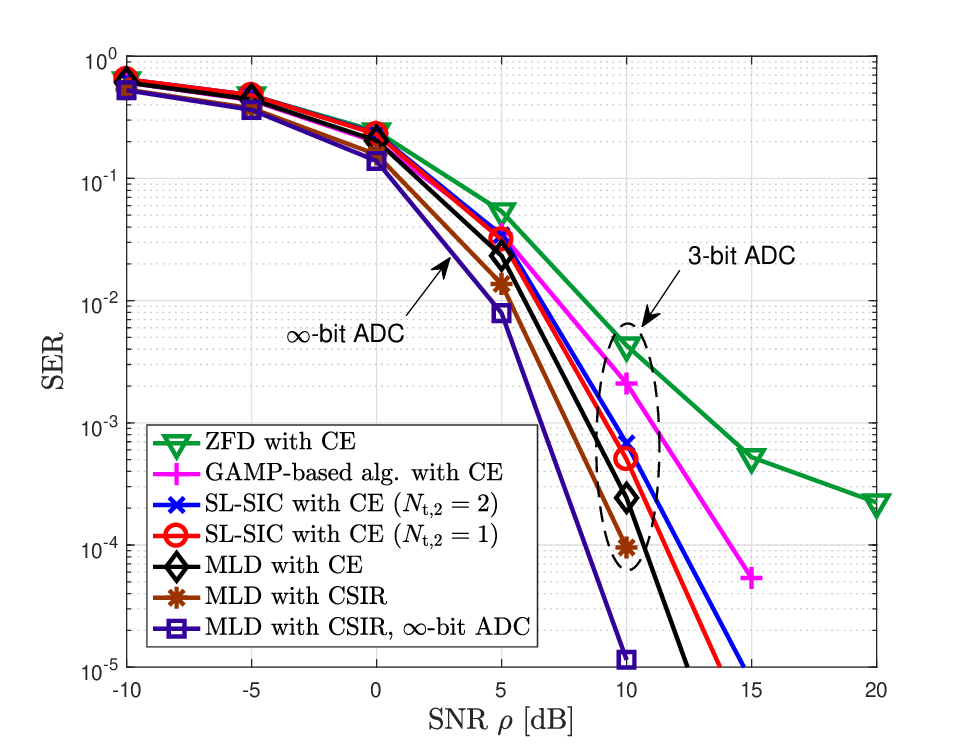, width=6.8cm}}
	\caption{SER vs. SNR of the proposed SL-SIC method and conventional detection methods in a MIMO system with low-resolution ADCs when 4-QAM modulation is employed with $N_{\rm t}=6$ and $N_{\rm r}=16$. For the detection methods with channel estimation (CE), the GAMP-based CE method in \cite{Wen2016} is adopted with pilot signals of length $T_{\rm t}\!=\!50$. For the SL-SIC method, the number of artificial training vectors is set by $L=20$.}\vspace{-3mm}
	\label{fig:SIC}
\end{figure*}

In Fig.~\ref{fig:SL}, we plot the SER of the framework proposed in Section~\ref{sec:SL} for two different cases: $T_{\rm t}\!=\!12$ (Fig.~\ref{fig:SL}(a)) and $T_{\rm t}\!=\!48$ (Fig.~\ref{fig:SL}(b)). These two cases correspond to $L\!=\!3$ and $L\!=\!12$, respectively, when adopting the efficient learning method developed in Section~\ref{sec:training}. The SER of the proposed framework is compared with that of MLD in \cite{Wang2014} and ZFD in \cite{Jacobsson2017}. Both methods are based on least-squares-based channel estimation (CE) method in \cite{Risi2016} with pilot signals of length $T_{\rm t}$. As a performance benchmark, we also plot the SER lower bound achieved by MLD with perfect CSIR. Fig~\ref{fig:SL} shows that in a high SNR regime, the proposed framework with MCD achieves the lowest SER regardless of SNRs and system parameters. In this case, the SER of MLD is severely degraded because it relies on an integral-form metric that is sensitive to the channel estimation error when the SNR is high. In a low-to-moderate SNR regime, the proposed framework with MCD shows a similar SER performance to MLD, but still outperforms ZFD. Among two detection methods (eMLD and MCD) developed for the proposed framework, MCD shows a better detection performance. The reason is that the performance of eMLD is degraded when the number of training repetitions, $L$, is not sufficiently large, while MCD is relatively robust to the value of $L$. It is also noticeable that MLD and MCD have the same order of detection complexity, because both methods compute the metrics of all possible symbol vectors, as shown in \cite{Wang2014} and \eqref{Eq_MCD}. The above results demonstrate that the proposed framework with MCD is an effective data detection method for MIMO systems with low-resolution ADCs, particularly when the number of possible symbol vectors $K\!=\!M^{N_{\rm t}}$ is comparable to the pilot length $T_{\rm t}$.

In Fig.~\ref{fig:SIC}, we plot the SER of the SL-SIC method proposed in Section~\ref{sec:SIC} for two different cases: $B\!=\!2$ (Fig.~\ref{fig:SIC}(a)) and $B\!=\!3$ (Fig.~\ref{fig:SIC}(b)), compared to the SERs of MLD in \cite{Wang2014}, the GAMP-based detection algorithm\footnote{In this method, we perform a joint channel-and-data estimation algorithm in \cite{Wen2016} for the given estimated channel matrix, by setting all signals as data signals (i.e., $T_{\rm t}\!=\!0$).} in \cite{Wen2016}, and ZFD in \cite{Jacobsson2017}. All detection methods are based on the GAMP-based CE method\footnote{In this method, we perform a joint channel-and-data estimation algorithm in \cite{Wen2016} for the given pilot signals, by setting all signals as pilot signals (i.e., $T_{\rm d}\!=\!0$).} in \cite{Wen2016} with pilot signals of length $T_{\rm t}\!=\!50$. As a performance benchmark, we also plot the SERs of MLD with perfect CSIR when low-resolution ADCs or infinite-bit ADCs is employed. Fig~\ref{fig:SIC} shows that the SER of MLD with CE is lower than the SER of the proposed method, but the difference between two SERs is small particularly when $B\!=\!3$. Meanwhile, the proposed method achieves a significant reduction in the computational complexity compared to MLD; the size of the search space for the proposed method are roughly $25\%$ and $6.6\%$ of that of MCD when $N_{{\rm t},2}\!=\!1$ and $N_{{\rm t},2} \!=\! 2$, respectively. Although the GAMP-based algorithm and ZFD may require less detection complexity than the proposed method does, their losses in the SER performance are considerable. Another important observation is that only the proposed method can adjust the tradeoff between the performance and the complexity, by using a design parameter $N_{{\rm t,2}}$. Therefore, the proposed SL-SIC method is useful to improve the detection performance-complexity tradeoff in a MIMO system with low-resolution ADCs.

\vspace{-1mm}
\subsection{Validation of Analysis}
We validate the analysis in Section~\ref{sec:analysis} by simulations for a MIMO system with the one-bit ADCs.

\begin{figure}[t]
	\centering
	\epsfig{file=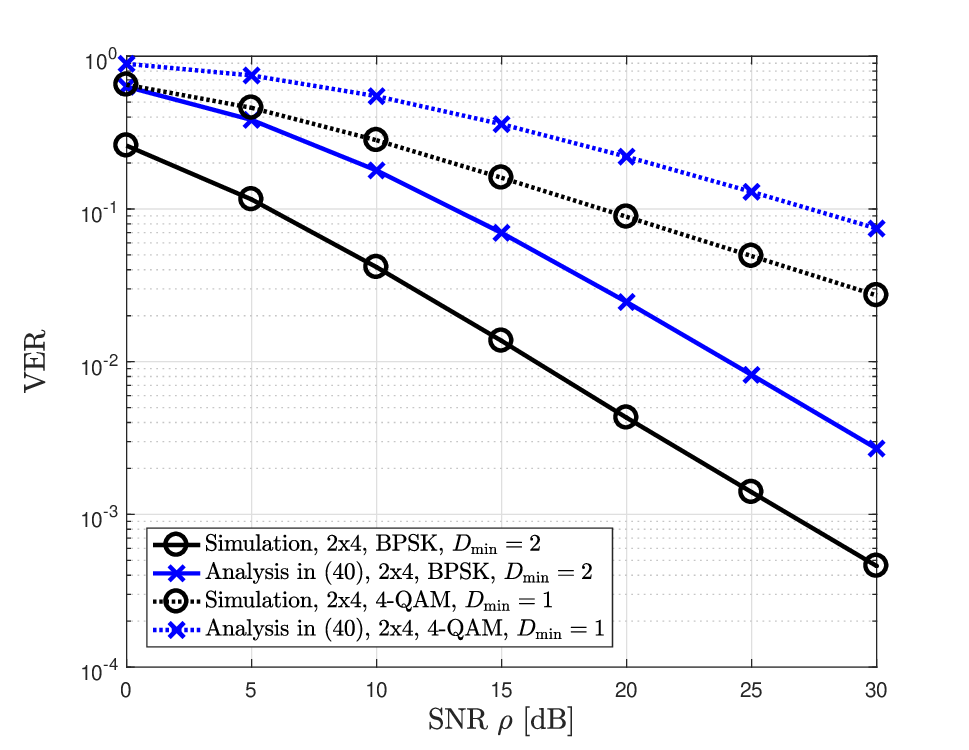, width=6.8cm}
	\caption{Comparison between analysis and simulation results for the vector error rate of the proposed framework when one-bit ADCs are employed with $N_{\rm t}=2$, $N_{\rm r}=4$, and BPSK or 4-QAM modulation. We use Monte-Carlo simulations with 5000 random generations of Rayleigh-fading channels and average out the results separately for each value of $D_{\rm min}$.}\vspace{-3mm}
	\label{fig:sim:Perr}
\end{figure}

Fig.~\ref{fig:sim:Perr} compares the upper bound of VER, derived in \eqref{Eq_Thm1}, with VER achieved by the proposed framework with MCD under the perfect learning assumption in \eqref{Eq_perf_train}. Fig.~\ref{fig:sim:Perr} shows that VER obtained by simulation is lower than the analyzed upper bound. Although the difference between the simulated VER and the analyzed upper bound is considerable due to the use of a loose upper bound in \eqref{Eq_V-A_Pei2}, the analyzed upper bound shows a similar VER slope to the simulation results  for every SNR value; thereby, this result is still useful to characterize the VER achieved by the proposed framework. Another important observation is that both the simulated VER and the analyzed upper bound are lower for $D_{\rm min}=2$ than for $D_{\rm min}=1$. These results coincide with the analysis in Section~\ref{sec:analysis:VER}, which implies that the VER decreases as $D_{\rm min}$ increases.

\begin{figure}[t]
	\centering
	\epsfig{file=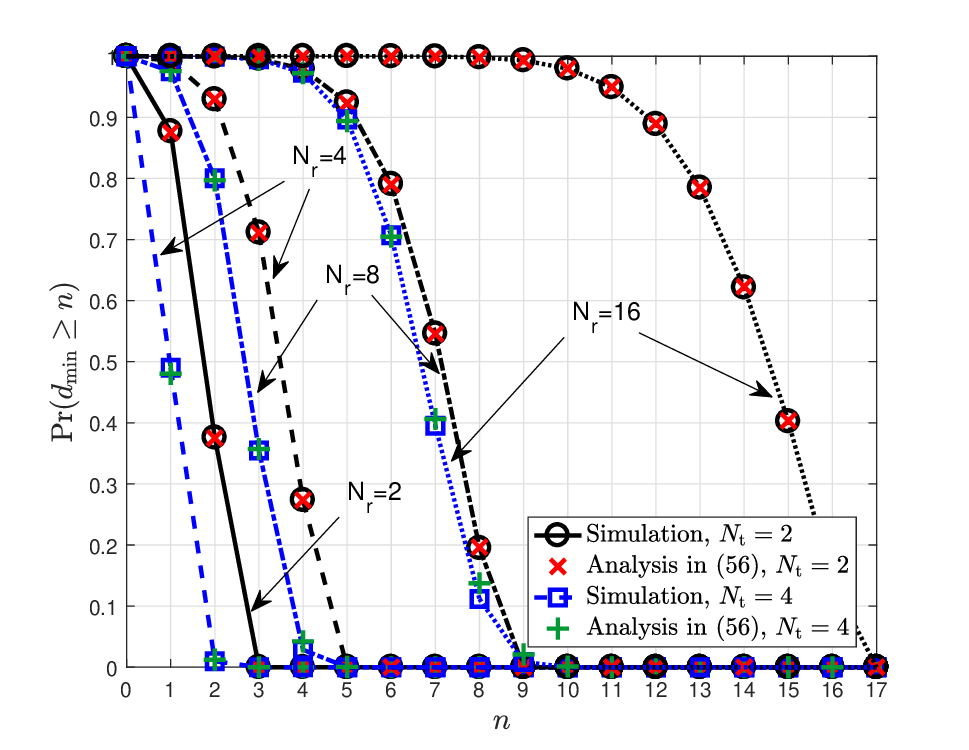, width=6.8cm}
	\caption{Comparison between analysis and simulation results for the CCDFs of $d_{\rm min}$ when BPSK modulation and one-bit ADCs are employed.}\vspace{-3mm}
	\label{Fig CCDF}
\end{figure}

Fig.~\ref{Fig CCDF} compares analysis and simulation results for the CCDFs of $d_{\rm min}$. For $N_{\rm t}=2$, the analyzed and simulated CCDFs are almost the same regardless of $N_{\rm r}$. For $N_{\rm t}=4$, although the simulated CCDF is not exactly the same as the analyzed CCDF, the difference between two CCDFs is negligible. These results validate the analysis given in Theorem~\ref{Thm2}. One important observation in Fig.~\ref{Fig CCDF} is that the value of $d_{\rm min}$ indeed increases as $N_{\rm r}$ increases; this result coincides with the intuition obtained from Corollary~2.

\vspace{-1mm}
\section{Conclusion}
In this paper, we have presented a novel communication framework for a MIMO system with low-resolution ADCs, inspired by supervised learning. Using this framework, we have shown that learning the nonlinear input-output system is an effective approach for the data detection. We have also revealed an interesting resemblance between the data detection problem in wireless communications and the classification problem in supervised learning. For the case of one-bit ADCs, we have analyzed the VER of the presented framework. The analysis results show that the upper bound of the VER decreases exponentially with the minimum distance that can increase with the number of receive antennas. Simulation results show that the presented framework is superior to conventional detection techniques that are based on channel estimation.

An important direction for future research is to extend the presented framework to frequency-selective channels. For this extension, some prior work in \cite{Mollen2017,Studer2016} can be jointly considered. Another interesting extension is to apply the presented framework to precoded MIMO systems that use low-resolution digital-to-analog converters (DACs) at a transmitter. This extension has a great potential to reduce a power consumption at the transmitter for downlink massive MIMO systems and/or wideband communication systems. It would also be interesting to optimize the detection rule of the presented framework by considering various kernel functions based on the empirical conditional PMFs.

\end{document}